\documentclass[12pt]{article}
\pdfoutput=1
\usepackage{makeidx}
\makeindex
\usepackage[a4paper]{geometry}
\usepackage{jheppub,amsmath,amssymb,amsfonts,amsxtra,mathrsfs,graphics,graphicx,amsthm,epsfig,ytableau,bm,longtable,float,color,tikz,mathtools,xfrac,footnote,rotating,lscape}
\usepackage{dsfont}
\usepackage{multicol}
\usepackage{lipsum}
\usepackage{textgreek}
\usetikzlibrary{decorations.pathmorphing}
\usetikzlibrary{decorations.markings}
\usetikzlibrary{quotes,arrows.meta}
\usetikzlibrary{arrows, decorations.markings, calc, fadings, decorations.pathreplacing, patterns, decorations.pathmorphing, positioning}
\usepackage{tikz-cd}

\usepackage{fixmath} 
\usepackage{scalerel}
\newlength\bshft
\def\fakebold#1{\ThisStyle{\ooalign{$\SavedStyle#1$\cr%
  \kern-\bshft$\SavedStyle#1$\cr%
  \kern\bshft$\SavedStyle#1$}}}

\usetikzlibrary{positioning,shapes}
\usetikzlibrary{chains}
\usetikzlibrary{arrows,fit,decorations.pathreplacing}
\tikzstyle{every picture}+=[remember picture]
\tikzstyle{na} = [baseline=-.5ex]

\usepackage{empheq}
\usepackage{multirow}
\usepackage{booktabs}
\usepackage[american]{babel}

\usepackage[latin1]{inputenc}

\makeatletter
\newcommand{\vast}{\bBigg@{1}}
\newcommand{\Vast}{\bBigg@{5}}
\makeatother

\usepackage{hyperref}

\numberwithin{equation}{section}

\newcommand{\cf}{\textit{cf.}}
\newcommand{\st}{\textit{s.t.}}
\newcommand{\eg}{\textit{e.g.}}

\newcommand{\ie}{\textit{i.e.}}

\newcommand{\ii}{\mathrm{i}}
\newcommand{\?}{\;\!}

\numberwithin{equation}{section}

\newcommand{\be}{\begin{equation}} \newcommand{\ee}{\end{equation}}
\newcommand{\bea}{\begin{equation} \begin{aligned}} \newcommand{\eea}{\end{aligned} \end{equation}}

\def\U{\mathrm{U}}
\def\SO{\mathrm{SO}}

\def\SU{\mathrm{SU}}

\def\USp{\mathrm{USp}}

\newcommand{\rd}{\mathrm{d}}

\newcommand{\Vol}{\mathrm{Vol}}

\DeclareMathOperator{\Tr}{Tr}

\DeclareMathOperator{\im}{\mathbb{I}m}

\newcommand{\cH}{\mathcal{H}}

\newcommand{\cK}{\mathcal{K}}

\newcommand{\cN}{\mathcal{N}}
\newcommand{\cO}{\mathcal{O}}

\newcommand{\cQ}{\mathcal{Q}}

\newcommand{\cV}{\mathcal{V}}

\newcommand{\bH}{\mathbb{H}}

\newcommand{\bZ}{\mathbb{Z}}

\usepackage[Symbol]{upgreek}
\usepackage{bm}

\usepackage{calligra}
\DeclareMathAlphabet{\mathcalligra}{T1}{calligra}{m}{n}

\theoremstyle{plain}

  \theoremstyle{definition}

\providecommand{\examplename}{Example}
\providecommand{\theoremname}{Theorem}

\makeatletter
\g@addto@macro\bfseries{\boldmath}
\makeatother

\makeatletter
\newcommand*{\rom}[1]{\expandafter\@slowromancap\romannumeral #1@}
\makeatother

\title{Supersymmetric R{\'e}nyi entropy and charged hyperbolic black holes}

\author[a]{Seyed Morteza Hosseini,}
\author[b,c,d]{Chiara Toldo,}
\author[e]{and Itamar Yaakov}
\affiliation[a]{Kavli IPMU (WPI), UTIAS, The University of Tokyo, Kashiwa, Chiba 277-8583, Japan}
\affiliation[b]{ Kavli Institute for Theoretical Physics, Kohn Hall, \\
	University of California Santa Barbara, CA, 93106}
\affiliation[c]{Centre de Physique Th\'eorique  (CPHT), Ecole Polytechnique,\\ 91128 Palaiseau Cedex, France}
\affiliation[d]{Institut de Physique Th\'eorique, Universit\'e Paris Saclay, CEA,  \\ CNRS, Orme des Merisiers, 91191 Gif-sur-Yvette Cedex, France}
\affiliation[e]{INFN - Sezione di Milano Bicocca \\ Dipartimento di Fisica, Edificio U2, Piazza della Scienza 3,
I-20126 Milano, MI, Italy}
\emailAdd{morteza.hosseini@ipmu.jp}
\emailAdd{chiara.toldo@polytechnique.edu}
\emailAdd{itamar.yaakov@mib.infn.it}

\preprint{IPMU19-0180, CPHT-045/11 2019}

\abstract{ The supersymmetric R{\'e}nyi entropy across a spherical entangling surface in a $d$-dimensional SCFT with flavor defects is equivalent to a supersymmetric partition function on $\mathbb{H}^{d-1} \times \mathbb{S}^1$, which can be computed exactly using localization. We consider the holographically dual BPS solutions in $(d +1)$-dimensional matter coupled supergravity $(d = 3 , 5)$, which are charged hyperbolically sliced AdS black holes. We compute the renormalized on-shell action and the holographic supersymmetric R{\'e}nyi entropy and show a perfect match with the field theory side. Our setup allows a direct map between the chemical potentials for the global symmetries of the field theories and those of the gravity solutions. We also discuss a simple case where angular momentum is added.}

\begin{document}

\setcounter{tocdepth}{2}
\maketitle

%
%

\date{Dated: \today}

\section{Introduction}
\label{sect:intro}

The entanglement entropy of the vacuum is an example of a universal
observable in quantum field theory, independent of the existence of
a particular set of fields, which has many interesting and useful properties. Most prominent among these are its
monotonicity properties as a function of the size of the entangling region \cite{Casini:2012ei},
and the existence of a simple geometric interpretation in the context
of holography \cite{Ryu:2006bv}. We refer the reader to the review \cite{Nishioka:2018khk} for more information.

The R{\'e}nyi entropy is a one parameter refinement of the entanglement
entropy. Besides containing additional information, the R{\'e}nyi entropy
is notable for having a straightforward Euclidean path integral interpretation
known as the replica trick \cite{Calabrese:2004eu}.
Supersymmetric R{\'e}nyi entropy (SRE) is a twisted version, in the sense of $(-1)^F$, of R{\'e}nyi
entropy which can be defined for supersymmetric theories in a variety
of spacetime dimensions and with varying amounts of supersymmetry
\cite{Nishioka:2013haa,Hama:2014iea,Huang:2014pda,Crossley:2014oea}.
Unlike R{\'e}nyi entropy, SRE can be calculated exactly at arbitrary
coupling using the method of supersymmetric localization. It nevertheless
shares many of the interesting properties of the untwisted version,
including the ability to recover the entanglement entropy as a limit. 

In a $d$-dimensional superconformal field theory (SCFT),
the SRE for a $d-2$-dimensional spherical entangling surface can be computed using the partition function on a $d$-sphere,
branched $n$ times over a maximal $d-2$-sphere, where the metric has a conical singularity. In holographically dual solutions, gravity becomes dynamical and the issue arises of how to treat such a singularity.
By conformally mapping the branched sphere to $\mathbb{H}^{d-1} \times \mathbb{S}^1$, where $\mathbb{H}$ denotes hyperbolic space, the singularity is pushed to infinity.  The R{\'e}nyi entropy is mapped to the thermal entropy in this space, with the new Euclidean time having periodicity $\beta = 2\pi n$.  The SRE is likewise mapped to a twisted thermal partition function. The details of the singularity are encoded in the boundary conditions on this space. The gravity duals are hyperbolically sliced solutions, so-called ``topological" black holes, whose boundary is indeed of the form $\mathbb{H}^{d-1} \times \mathbb{S}^1$. 

The computation of the SRE in $d$-dimensional models ($d=2,3,4,5,6$) with a holographic dual was performed, respectively, in \cite{Giveon:2015cgs,Mori:2015bro,Nishioka:2014mwa,Huang:2014gca,Crossley:2014oea,Hama:2014iea,Alday:2014fsa}. The matching with the gravity computation of the SRE was achieved with supergravity hyperbolic black holes supported by a single gauge field, which corresponds to the graviphoton. Here we take this one step further, by considering supergravity backgrounds with more general couplings, in particular vector multiplets. The corresponding dual field theory computation therefore includes fugacities for the global symmetries of the theory, equivalently co-dimension two flavor vortex defects in the $d$ sphere picture. In gravity, we work with four and six-dimensional supergravity solutions, achieving a match with the field theory SRE in $d=3,5$ by evaluating the supergravity renormalized on-shell action. We choose to work with $d$ odd because the finite part of the free energy in the field theory is believed to be universal. For comparison, in the even $d$ case the coefficient of the Weyl anomaly is always universal, while the subleading piece may only be universal in the presence of a sufficient amount of supersymmetry \cite{Gerchkovitz:2014gta}. By working in even-dimensional supergravity, we also avoid subtleties in holographic renormalization schemes related to the Casimir energy, see \eg\;\cite{Genolini:2016sxe,Papadimitriou:2017kzw,An:2017ihs}. Let us however mention that the SRE of supergravity solutions in $d + 1 = 5, 7$, coupled to matter were compared to the field theory result, respectively, in \cite{Huang:2014pda,Yankielowicz:2017xkf}.

The aim of this paper is twofold. On one hand, we wish to investigate how the SRE is computed holographically in the case where matter couplings are incorporated -- in the present case,
this consists in considering hyperbolic black hole solutions supported by vector multiplets. On the other hand, our setup allows to directly map the fugacities appearing in the field theory computation to the black hole chemical potentials.
The mapping that we obtain is then rather manifest.%
\footnote{For instance, in the case of rotating electric black holes, an elegant prescription to map the black hole chemical potentials to the field theory ones was recently put forward in \cite{Cabo-Bizet:2018ehj}. This procedure requires taking an extremal limit of a family of supersymmetric, complexified solutions, and the definition of the black hole chemical potentials via appropriate subtraction of the extremal BPS values. In our framework, upon Wick-rotating the BPS black hole solution we are left with a regular geometry with topology $\mathbb{R}^2 \times \bH^{d-1}$, where  a formal finite temperature can be defined. This allows us to directly map the chemical potentials in gravity into those on the field theory side, with no need for such a subtraction.} 

The paper is organized as follows. We will first provide results for the supersymmetric R{\'e}nyi entropy with flavor fugacities for specific models: the ABJM model in $d=3$, and a $\mathcal{N}=1$, $\USp(2N)$ gauge theory with $N_f$ fundamental and one anti-symmetric hypermultiplets in $d=5$. These models have well known gravity dual descriptions. We then focus on the gravity duals to SRE in four and six dimensions, which are hyperbolic black holes. We spell out the solutions, which are new in the $d = 6$ case, and compute their renormalized on-shell action. We show that this matches with the SRE computation. In appendix \ref{AppA}, we explicitly construct the Killing spinors for the hyperbolic black holes. Appendix \ref{AppB} shows the computation of the renormalized on-shell action via holographic renormalization techniques and appendix \ref{AppC} shows that the black hole charges computed from supergravity match those computed in the SCFT. In appendix \ref{AppD}, we present a simple example of a rotating hyperbolic black hole which generalizes the static case in section \ref{warmup}, and provide the value of its renormalized on-shell action.

\section{Field theory}
\label{sect:field_theory}

In this section we calculate the free energy of SCFTs on $\bH^{d-1} \times \mathbb{S}^{1}$ that are holographically dual to our hyperbolic BPS black holes.
We first introduce the supersymmetric R{\'e}nyi entropy (SRE) and its deformation by BPS vortex defects.
We then describe the relationship of these defects to black hole chemical potentials. 
Using supersymmetric localization, we construct an appropriate matrix model which captures the exact answer for the free energy.
Finally, we use large $N$ techniques to explicitly evaluate the matrix model for field theories dual to the black hole solutions.
 
\subsection{Supersymmetric R{\'e}nyi entropy}
\label{subsect:SRE}

We briefly review the definition of R{\'e}nyi entropy and its supersymmetric counterpart (SRE).
We then show how co-dimension two defect operators alter the localization result for SRE.
Finally, we relate such defects to chemical potentials in the partition
function on hyperbolic space. 

\subsubsection{Definition of R{\'e}nyi entropy}

Following the notation in \cite{Nishioka:2018khk},
we define entanglement entropy for a vacuum state $\Psi$ by first making
a choice of a subregion $A$ of a spatial slice. The complement will
be denoted by $B=\bar{A}$. We make the assumption that the Hilbert space
of the theory can be likewise locally split as%
\footnote{For a critical discussion of the validity of this assumption, see references
in footnote $3$ of \cite{Nishioka:2018khk}. The subtleties associated
with this splitting will not affect our results.}
\be
\mathcal{H}=\mathcal{H}_{A}\otimes\mathcal{H}_{B} \, .
\ee
We then form the reduced density matrix corresponding to $A$
\be
\rho_{A}\equiv\text{tr}_{B}\left|\Psi\right\rangle \left\langle \Psi\right| \, .
\ee
The entanglement
entropy associated to $A$ can be defined as the von Neumann entropy
of $\rho_{A}$,
\be
S\left(A\right)\equiv- \Tr \rho_{A}\log\rho_{A} \, .
\ee
The R{\'e}nyi entropy is a one parameter refinement of the entanglement
entropy defined by 
\be
S_{n}\left(A\right)\equiv\frac{1}{1-n}\log\text{tr}\rho_{A}^{n}\, ,\quad n\in\mathbb{N}\, .
\ee
It satisfies the relation%
\be
\lim_{n\rightarrow 1}S_{n}\left(A\right)=S\left(A\right),
\ee
where the limit is understood to be taken using an appropriate continuation to non-integer $n$.
We will restrict our attention to the case where $A$ is the $d-1$ ball and the entangling surface is $\partial A=\mathbb{S}^{d-2}$.

The R{\'e}nyi entropy of a quantum field theory is, in general, divergent. However, for $d$
odd the finite part of the R{\'e}nyi entropy of a CFT
is believed to be a universal observable (see \cite{Nishioka:2018khk} and references within). 

The R{\'e}nyi entropy can alternatively be computed using the replica trick \cite{Calabrese:2004eu}. One considers the path integral on an
$n$-fold cover of the original spacetime branched around the entangling
surface $\partial A$. Denoting the partition function on this space by $Z_n$, we will \emph{define}
the $n$-th R{\'e}nyi entropy for a positive integer $n$ by%
\footnote{The absolute value, which is absent from the usual definition, is used here to avoid some subtleties associated with possible non-universal terms in the SRE defined later on. See \cite{Nishioka:2013haa} for a discussion of the $d=3$ case.}
\begin{align}
S_{n}\equiv\frac{1}{1-n}\log\left|\frac{Z_{n}}{\left(Z_{1}\right){}^{n}}\right|\, .\label{Renyi_Definition}
\end{align}
This definition is incomplete because the branching means that the
spacetime corresponding to $Z_{n}$ is not smooth but has conical singularities. One could complete
the definition by specifying appropriate boundary conditions for all
fields at $\partial A$. We will instead concentrate on the definition
of SRE, reviewed in section \ref{subsect:SRE}, which uses a particular prescription for smoothing out the singularities \cite{Nishioka:2013haa}. 

The line element on a branched $d$ sphere is defined as the round sphere metric with a different coordinate range 
\bea
 \rd s^{2} & =\ell^{2}\left(\rd \theta^{2}+\sin^{2}(\theta) \rd \tau^{2}+\cos^{2}(\theta) \rd s_{\mathbb{S}^{d-2}}^{2}\right) \, , \\
 &\theta\in\left[0,\pi/2\right],\quad\tau\in\left[0,2\pi n\right) .
\eea
This metric has a conical singularity along the co-dimension two maximal
$d-2$ sphere at $\theta=0$. 

For $n$ a positive integer, the branched sphere is related by a Weyl
transformation to the branched version of $\mathbb{R}^{d}$ used to
define the $n$-th R{\'e}nyi entropy \cite{Casini:2011kv}. 
In order to avoid working with a singular space, we can conformally map this space by $\cot (\theta) = \sinh (\chi)$ to  $\mathbb{H}^{d-1}\times\mathbb{S}^{1}$ with line element
\bea \label{hypn}
 \rd s^2_{\mathbb{H}^{d-1} \times \mathbb{S}^{1}} & = \rd \tau^2 + \rd \chi^2 + \sinh(\chi)^2 \rd s^2_{\mathbb{S}^{d-2}} \, , \\
 &\chi\in\left[0, \infty \right],\quad\tau\in\left[0,2\pi n\right) .
\eea
The R{\'e}nyi entropy maps to the thermal entropy in this space with inverse temperature $\beta=2\pi n$. The singularity at $\theta=0$ is mapped to $\chi \rightarrow \infty$ \cite{Casini:2011kv}. 

\subsubsection{Definition of supersymmetric R{\'e}nyi entropy}

The supersymmetric R{\'e}nyi entropy (SRE) is a twisted version, in the sense of $(-1)^F$, of R{\'e}nyi entropy \cite{Nishioka:2013haa,Hama:2014iea,Huang:2014pda,Crossley:2014oea}.
In order to preserve supersymmetry in SRE, one must give nonzero values
to additional fields, aside form the metric, in the background supergravity multiplet to which
the SCFT is coupled \cite{Festuccia:2011ws,Closset2012,Dumitrescu:2012at,Dumitrescu:2012ha}. 
Specifically, one needs to turn
on a background R-symmetry gauge field, $A^{(R)}$,
which is flat in the bulk of the space and has a delta function like
field strength supported on the singularity \cite{Nishioka:2013haa}. 
For example, in a three-dimensional $\mathcal{N}=2$ field theory we have \cite{Nishioka:2013haa}%
\footnote{The sign of $A^{(R)}$ chosen here, which is correlated with the choice of Killing spinor preserved by SRE, corresponds to our gravity conventions and is opposite to the one chosen in \cite{Nishioka:2013haa}.}
\be \label{Rsymm}
A^{(R)}=-\frac{n-1}{2n}\rd\tau \, .
\ee
After the additional Weyl transformation to $\mathbb{H}^{d-1}\times\mathbb{S}^{1}$, the SRE is related to a twisted, in the sense of $(-1)^F$, version of the thermal partition function which we can
call the \emph{hyperbolic index}, in analogy with the superconformal index \cite{Kinney:2005ej,Romelsberger:2005eg}.
A representation of this quantity as a trace over the Hilbert space $\cH_{\bH^{d-1}}$ of states on $\mathbb{H}^{d-1}$ was given in \cite{Zhou:2016kcz}. Including flavor charges, we can write\footnote{As an index, $Z^{\text{susy}}_{n}$ does not change under renormalization group flow, and thus can be computed either in the UV or the IR SCFT. The parameter $n$ is a chemical potential for a combination of charges commuting with the supercharge, similar to those found in \cite{Kinney:2005ej,Romelsberger:2005eg}.} 
\be
\label{trace_representation} Z^{\text{susy}}_{n} = \Tr_{\cH_{\bH^{d-1}}} e^{-2\pi n\left(H-\ii \sum_I \alpha^I Q_{I}^{\text{flavor}}+\ii\frac{n-1}{n}Q_R\right)} \, ,
\ee
where $H$ is the Hamiltonian, $Q_R$ is the R-symmetry charge, $Q_{I}^{\text{flavor}}$ are flavor charges, and the $\alpha^I$ are flavor chemical potentials. The SRE is then defined as
\be
 \label{def:SRE}
S^\text{SRE}_n \equiv \frac{1}{1-n}\log \frac{Z^{\text{susy}}_{n}}{\left(Z^{\text{susy}}_{1}\right)^n}\,.
\ee

\subsubsection{Localization and deformation of SRE}

The partition function defining the SRE can be computed exactly using
the method of supersymmetric localization \cite{Witten:1988ze,Pestun:2007rz}.
In the case of SRE in three dimensions, the matrix model one gets from
localization coincides with the one used to compute the partition
function on the squashed sphere with the squashing parameter related to $n$ in a simple way \cite{Hama:2011ea,Nishioka:2013haa}.%
\footnote{This is true at the level of the matrix model, not just the final result.}
This relationship continues to hold for higher dimensions and we consequently make no distinction between the free energy in the two matrix models.

The partition function defining the SRE can be refined by supersymmetric
deformations while remaining amenable to localization \cite{Kapustin:2009kz,Nishioka:2013haa}.%
\footnote{We describe the situation in three dimensions. The situation in five dimensions is analogous.} Deformations include masses
for matter multiplets and Fayet-Iliopoulos (FI) terms for abelian vector multiplets. These deformations break conformal invariance. Additionally, the form of
the coupling of the theory to the background supergravity fields,
including $A^{(R)}$, depends on a choice of R-symmetry
current. If the R-symmetry is abelian, one may choose an arbitrary
linear combination of R-symmetry and abelian flavor symmetry currents. In
an SCFT, a particular combination, the result of dynamical mixing, is dictated by the superconformal algebra
where the R-symmetry transformations appear \cite{Jafferis:2010un,Closset:2012vg}.

Supersymmetric operators can also be added to the SRE. These include Wilson loops and co-dimension two vortex defects \cite{Kapustin:2009kz,Kapustin:2012iw,Drukker:2012sr}. The latter are inserted by demanding that the fields in the path integral
have prescribed singularities on the defect worldvolume \cite{Gukov:2014gja}.
If the defect is in a flavor symmetry this is equivalent to introducing
background flavor symmetry gauge fields which are flat outside the defect.
In fact, the deformation leading from the usual sphere partition function to the SRE is itself such a defect, embedded in the background supergravity multiplet. 
Due to this, addition of flavor defects to the SRE, oriented
along the same sub-manifold, is essentially the same as the R-symmetry
mixing effect described above. However, the strength of the defect
is now unrelated to the superconformal algebra and represents a deformation
of the SRE. 
In the hyperbolic picture, such a defect is mapped to the holonomy of a flavor symmetry connection
along the time direction, \ie\;a flavor fugacity. The chemical potentials $\alpha$ for such a fugacity are linearly related to the $A^\text{flavor}_{\tau}$ flavor gauge fields introduced below, with a proportionality constant which depends on the normalization of the charges.

\subsubsection{The SRE matrix model deformed by defects \label{def}}

The matrix model for the round sphere deformed by co-dimension two
defects, in dimensions $d = 3, 4, 5$ was derived in \cite{Nishioka2016}.
It was shown that a background $\U(1)$ flavor symmetry connection
$A^\text{flavor}$ with holonomy $\exp\left(2\pi \ii A^\text{flavor}_{\tau}\right)$ induces,
after localizing to a matrix model, a mass deformation term%
\be
m_{\text{defect}}=- \ii A^\text{flavor}_{\tau} \, .
\ee
The fact that the mass is imaginary is part of the relationship to
R-symmetry mixing. 
The large $N$ limit of the same matrix models in the presence of R-symmetry mixing or of mass terms has previously been derived in \cite{Martelli:2011fu,Imamura:2011wg,Chang:2017mxc}. The mixing parameters are
usually called $\Delta$,%
while masses are denoted by $m$. Besides
being purely imaginary, the mass term induced by the defect also has an
origin which is naturally $A^\text{flavor}_{\tau}=0$. This is true also for the
real ``physical masses'' $m$. On the other hand, in a theory which
has a non-abelian R-symmetry group, the $\Delta$'s have an origin
which is determined by the canonical R-charge, or the canonical dimensions,
of matter multiplets. In three dimensions this is $\Delta=1/2$,
while in five dimensions it is $\Delta=3/2$.

Taking all this into account, and using the relationship between $n$
and the squashing parameter $b$ derived in \cite{Nishioka2016},
the defect deformed three-dimensional matrix models are given by those of \cite{Martelli:2011fu} with the substitution%
\footnote{The setup is symmetric with respect to inversion of $b$. In order to conform to the notation in \cite{Martelli:2011fu}, we set $b=1/\sqrt{n}$ instead of $b=\sqrt{n}$ as in \cite{Nishioka2016}.}
\be \label{map3d}
 \Delta_{\text{there}} = \frac{1}{2}+\frac{2n A^\text{flavor}_{\tau}}{n+1} \, , \qquad b_{\text{there}} = \frac{1}{\sqrt{n}} \, .
\ee
For five-dimensional $\cN=1$ theories appearing in \cite{Chang:2017mxc}, we can simply take
\be \label{6d_mapping}
 m_{\text{there}} = - \ii A^\text{flavor}_{\tau} \, , \qquad
 \vec{\omega}_{\text{there}} = \left(1,1, 1 / n \right) \, .
\ee

We will adopt a democratic convention for the deformation parameters $\Delta$, whereby the physical parameters are augmented by one additional parameter and a constraint is imposed. 
Interpreting $\Delta$ as the result of a flavor defect, we will add a corresponding $A^\text{flavor}$. The constraint in terms of $A^\text{flavor}$ is simply
\be 
\label{chemical_potential_constraint} \sum_I A^{\text{flavor},I} = 0 \, .
\ee

\subsection[Squashed \texorpdfstring{$\mathbb{S}^3$}{S**3} free energy]{Squashed $\mathbb{S}^3$ free energy}
\label{sec:logZ:3D}

In this section, we review the squashed $\mathbb{S}^3$ partition function and its large $N$ limit, as analyzed in \cite{Martelli:2011fu,Imamura:2011wg}.
For the purpose of this paper, we consider the ABJM model \cite{Aharony:2008ug}, which is holographically dual to an AdS$_4 \times \mathbb{S}^7/\bZ_k$ background of M-theory.
ABJM is a three-dimensional $\mathcal{N}=6$ supersymmetric Chern-Simons-matter theory with gauge group $\U(N)_k \times \U(N)_{-k}$
(the subscripts represent the CS levels) with two pairs of bi-fundamental chiral fields $A_i$ and $B_i$, $i=1,2$, in the representation $({\bf N},\overline{{\bf N}})$ and $(\overline{{\bf N}},{\bf N})$ of the gauge group, respectively.
The chiral fields interact through the quartic superpotential
\be
 \label{superpotential:ABJM}
 W = \Tr \big( A_1B_1A_2B_2 - A_1B_2A_2B_1 \big) \, .
\ee

In the $\mathcal{N}=2$ formulation, the ABJM model has a $\U(2)\times \U(2)$ action which acts separately on the chiral fields $A_{1,2}$ and $B_{1,2}$. There is a $\U(1)^3$ subgroup of the Cartan of this group which preserves the superpotential, a particular linear combination of which is gauged. In addition, there are two topological $\U(1)_J$ symmetries. The current for one of these topological symmetries is set to zero by the equations of motion. Due to the appearance of Chern-Simons terms, the action of the other $\U(1)_J$ is mixed with the gauge group action. We will work in a gauge in which the fugacity conjugate to the remaining topological symmetry, which could be explicitly added using an FI parameter, is fixed to $1$. The remaining global symmetry group, which we will call the flavor group, is given by the $\U(1)^3$ compatible with the superpotential acting on the chiral fields. The model admits therefore a three-parameter space of flavor symmetry, or $\Delta$ type, deformations.%
\footnote{We would like to thank Alberto Zaffaroni for explaining this point.}

We introduce the R-charges $\Delta_I$, $I=1,\ldots,4$, one for each of the four fields $\{A_i,B_i\}$, satisfying
\be
 \label{constraint:ABJM}
 \sum_{I = 1}^{4} \Delta_I = 2 \, .
\ee
The partition function can be written as
\be
 Z_{\mathbb{S}^3_b} = \int_{- \infty}^{\infty} \left[ \prod_{i = 1}^{N} \frac{\rd \lambda_i}{2 \pi} \frac{\rd \tilde \lambda_i}{2 \pi} \right] e^{- F_{\mathbb{S}^3_b} (\lambda_i , \tilde \lambda_i)} \, ,
\ee
where
\bea
 \label{free_energy:S^3}
 F_{\mathbb{S}^3_b} & = 2 \log N! - \frac{i k}{4 \pi b^2} \sum_{i = 1}^{N} \Big( \lambda_i^2 - \tilde \lambda_i^2 \Big) \\
 & - \sum_{i < j}^{N} \left\{ \log \left[ 2 \sinh \left( \frac{\lambda_i - \lambda_j}{2} \right) \right] + \log \left[ 2 \sinh \left( \frac{\lambda_i - \lambda_j}{2 b^2} \right) \right] \right\} \\
 & - \sum_{i < j}^{N} \left\{ \log \left[ 2 \sinh \bigg( \frac{\tilde \lambda_i - \tilde \lambda_j}{2} \bigg) \right] + \log \left[ 2 \sinh \bigg( \frac{\tilde \lambda_i - \tilde \lambda_j}{2 b^2} \bigg) \right] \right\} \\
 & - \sum_{i,j=1}^{N} \sum_{a=1}^{2} S_2 \left( \frac{i \mathfrak{Q}}{2} (1 - \Delta_a ) - \frac{1}{2 \pi b} ( \lambda_i - \tilde \lambda_j ) \bigg| b \right) \\
 & - \sum_{i,j=1}^{N} \sum_{b=3}^{4} S_2 \left( \frac{i \mathfrak{Q}}{2} (1 - \Delta_b ) + \frac{1}{2 \pi b} ( \lambda_i - \tilde \lambda_j ) \bigg| b \right) \, .
\eea
Here, $\mathfrak{Q} =b + 1/b$ and $S_2 (\lambda | b)$ is the double sine function.

\paragraph*{Large $N$ free energy.}

Consider the following ansatz for the large $N$ saddle point eigenvalue distribution,
\be
 \label{Ansatz:largeN:3D}
 \lambda_j = N^{1/2} t_j + \ii v_j \, , \qquad  \tilde \lambda_j = N^{1/2} t_j + \ii \tilde v_j \, .
\ee
In the large $N$ limit, we define the continuous functions $t_j = t ( j / N )$ and $v_j = v (j / N)$, $\tilde v_j = \tilde v (j / N)$; and we introduce the density of eigenvalues
\be
 \label{rho(t)}
 \rho (t) = \frac{1}{N} \frac{\rd j}{\rd t} \, , \quad \st \int \rd t \rho (t) = 1 \, .
\ee
At large $N$ the sums over $N$ become Riemann integrals, for example,
\be
 \sum_{j = 1}^{N} \to N \int \rd t \rho (t) \, .
\ee
The large $N$ free energy is then given by \cite{Martelli:2011fu,Imamura:2011wg}
\bea
 \label{S^3:free_energy:functional}
 \frac{F_{\mathbb{S}^3_b} \left[ \rho(t) , \delta v(t) , \Delta_I | b \right]}{N^{3/2}} & = \frac{k}{2 \pi b^2} \int \rd t \rho(t) t \delta v(t) - \gamma \left( \int \rd t \rho(t) - 1 \right) \\
 & - \frac{b \mathfrak{Q}^3}{16} \sum_{a=1}^2 (2 - \Delta_a^+) \int \rd t \rho(t)^2 \left[ \left( 2 \frac{\delta v(t)}{b \mathfrak{Q}} + \pi \Delta_a^- \right)^2 - \frac{\pi^2}{3} \Delta_a^+ ( 4 - \Delta_a^+ ) \right] ,
\eea
where we defined $\delta v(t) \equiv v(t) - \tilde v(t)$, $\Delta_1^\pm \equiv \Delta_1 \pm \Delta_4$, $\Delta_2^\pm \equiv \Delta_2 \pm \Delta_3$, and we added the Lagrange multiplier $\gamma$ for the normalization of $\rho(t)$.
Setting to zero the variation of \eqref{S^3:free_energy:functional} with respect to $\rho(t)$ and $\delta v(t)$ we obtain the following saddle point configuration.
We have a central region where 
\be
 \begin{aligned}
 \rho (t)&= \frac{16 b \gamma + 4 \mathfrak{Q} k t (\Delta_1 \Delta_2-\Delta_3 \Delta_4 )}{4 \pi^2 b^2 \mathfrak{Q}^3 (\Delta_1+\Delta_3 ) (\Delta_2+\Delta_3 ) (\Delta_1+\Delta_4 ) (\Delta_2+\Delta_4 )} \, , \\[.5em]
 \delta v (t)&= \frac{2 \pi b \mathfrak{Q}^2 k t \sum_{a<b<c} \Delta_a \Delta_b \Delta_c - 4 \pi b^2 \mathfrak{Q} \gamma (\Delta_1 \Delta_2-\Delta_3 \Delta_4 )}{8 b \gamma + 2 \mathfrak{Q} k t (\Delta_1 \Delta_2 -\Delta_3 \Delta_4 )} \, ,
 \end{aligned}
 \qquad - \frac{2 b \gamma }{\mathfrak{Q} k \Delta_1}  < t < \frac{2 b \gamma }{\mathfrak{Q} k \Delta_3} \, .
\ee
When $\delta v = - \pi b \mathfrak{Q} \Delta_2$ on the left the solution reads
\be
\rho (t)= \frac{2 b \gamma + \mathfrak{Q} k t \Delta_2}{\pi^2 b^2 \mathfrak{Q}^3 (\Delta_1-\Delta_2 ) (\Delta_2+\Delta_3 ) (\Delta_2+\Delta_4 )} \, , \qquad  - \frac{2 b \gamma }{\mathfrak{Q} k \Delta_2} < t < - \frac{2 b \gamma }{\mathfrak{Q} k \Delta_1} \, ,
\ee
while when  $\delta v = \pi b \mathfrak{Q} \Delta_4$ on the right the solution is given by
\be
\rho (t) = - \frac{2 b \gamma - \mathfrak{Q} k t \Delta_4}{\pi^2 b^2 \mathfrak{Q}^3(\Delta_1+\Delta_4 ) (\Delta_2+\Delta_4 ) (\Delta_4-\Delta_3 )} \, , \qquad  \frac{2 b \gamma }{\mathfrak{Q} k \Delta_3} < t < \frac{2 b \gamma }{\mathfrak{Q} k \Delta_4} \, .
\ee
The normalization of $\rho(t)$ fixes the value of $\gamma$ as
\be
 \gamma = \frac{\pi \mathfrak{Q}^2}{\sqrt{2}} \sqrt{k \Delta_1 \Delta_2 \Delta_3 \Delta_4} \, .
\ee
Plugging the above solution back into \eqref{S^3:free_energy:functional} we obtain the squashed $\mathbb{S}^3$ free energy%
\footnote{The first equality arises from a virial theorem for the free energy \eqref{S^3:free_energy:functional}.}
\be \label{br_sph}
 F_{\mathbb{S}^3_b} (\Delta_I|\mathfrak{Q}) = \frac{2 N^{3/2}}{3} \gamma= \frac{\pi N^{3/2} \mathfrak{Q}^2}{3} \sqrt{2 k \Delta_1 \Delta_2 \Delta_3 \Delta_4} = \frac{\mathfrak{Q}^2}{4} F_{\mathbb{S}^3}  (\Delta_I) \, ,
\ee
where $F_{\mathbb{S}^3}$ is the free energy of ABJM on the round $\mathbb{S}^3$, \ie\;$b=1$, see \cite[sect.\,5]{Jafferis:2011zi}.
This is precisely \cite[(3.38)]{Martelli:2011fu}.

\subsection[Squashed \texorpdfstring{$\mathbb{S}^5$}{S**5} free energy]{Squashed $\mathbb{S}^5$ free energy}
\label{sec:logZ:5D}

In this section we review the large $N$ limit of the squashed $\mathbb{S}^5$ free energy of the $\USp(2N)$ gauge theory
with $N_f$ hypermultiplets in the fundamental representation and one hypermultiplet in the antisymmetric representation of $\USp(2N)$,  as analyzed in \cite{Chang:2017mxc}.
The gauge theories of interest live on the intersection of $N$ D4-branes and $N_f$ D8-branes and orientifold planes in type I' string theory and
are holographically dual to a warped AdS$_6 \times \mathbb{S}^4$ background of massive type IIA supergravity \cite{Intriligator:1997pq} (see also \cite{Brandhuber:1999np,Bergman:2012kr,Morrison:1996xf,Seiberg:1996bd}).

The perturbative partition function can be written as%
\footnote{We will neglect instanton contributions as they are exponentially suppressed in the large $N$ limit.}
\be
 Z_{\mathbb{S}^5_\omega}^{\text{pert}} = \int_{- \infty}^{\infty} \left[ \prod_{i = 1}^{N} \frac{\rd \lambda_i}{2 \pi} \right] e^{- F_{\mathbb{S}^5_\omega} (\lambda_i)} \, ,
\ee
where
\bea
 \label{free_energy:S^5}
 F_{\mathbb{S}^5_\omega} & = N \log 2 + \log N! - N \log S'_{3} ( 0 | \vec{\omega}) + (N-1) \log S_3 \left( \ii  m_{a} + \frac{\omega_{\text{tot}}}{2} \Big| \vec{\omega} \right) \\
 & + \frac{1}{\omega_1 \omega_2 \omega_3} \frac{4 \pi^3}{g_{\text{YM}}^2} \sum_{i = 1}^N \lambda_i^2 - \sum_{i > j}^{N} \log S_3 \left( \ii  \left[ \pm \lambda_i \pm \lambda_j \right] | \vec{\omega} \right)
 - \sum_{i = 1}^N \log S_3 \left( \pm 2 \ii  \lambda_i | \vec{\omega} \right) \\
 & + \sum_{i > j}^{N} \log S_3 \left( \ii  \left[ \pm \lambda_i \pm \lambda_j \right] + \ii  m_a + \frac{\omega_{\text{tot}}}{2} \Big| \vec{\omega} \right)
 + N_f \sum_{i = 1}^N \log S_3 \left( \pm \ii  \lambda_i + \ii  m_f + \frac{\omega_{\text{tot}}}{2} \Big| \vec{\omega} \right) \, ,
\eea
with $S_3 ( \lambda | \vec{\omega})$ being the triple sine function.
Here, $m_a$ and $m_f$ are the masses for the hypermultiplets in the antisymmetric and fundamental representations of $\USp(2N)$, respectively.
We also introduced the notation
\be
 \omega_{\text{tot}} \equiv \omega_1 + \omega_2 + \omega_3 \, , \qquad S_{3} (\pm z | \vec{\omega} ) \equiv S_3 (z | \vec{\omega}) S_3 (- z | \vec{\omega}) \, .
\ee

\paragraph*{Large $N$ free energy.}

We may restrict to $\lambda_i \geq 0$ due to the Weyl reflections of the $\USp(2N)$ group.
Consider the following ansatz for the large $N$ saddle point eigenvalue distribution,
\be
 \label{Ansatz:largeN:5D}
 \lambda_j = N^{\alpha} t_j \, ,
\ee
where $\alpha \in ( 0 , 1)$ will be determined later.
As in the previous section, at large $N$, we define the continuous function $t_j = t (j / N)$ and we introduce the density of eigenvalues $\rho(t)$, see \eqref{rho(t)}.
In the large $N$ limit, $\lambda_{i} = \cO(N^{1/2})$ (see \eqref{Ansatz:largeN:5D} with $\alpha = 1/2$).
Therefore, at large $N$, the contributions with nontrivial instanton numbers are exponentially suppressed.
In the continuum limit, the free energy \eqref{free_energy:S^5} is given by \cite{Chang:2017mxc}%
\footnote{Notice, that the free energy at large $N$ does \emph{not} depend on the masses of the $N_f$ fundamental hypermultiplets. As it was shown in \cite[(3.22)]{Chang:2017mxc} their contribution to the large $N$ free energy is of order $\cO(N^{3/2})$ and, thus, subleading.}
\bea
 \label{S^5:free_energy:functional}
 F_{\mathbb{S}^5_\omega} \left[ \rho(t) , m_a | \vec{\omega} \right] & = \frac{N^{1 + 3 \alpha}}{\omega_1 \omega_2 \omega_3} \frac{\pi ( 8 - N_f )}{3} \int_{0}^{t_*} \rd t \rho(t) | t |^3 - \mu \left( \int_{0}^{t_*} \rd t \rho(t) - 1 \right) \\
 & - \frac{N^{2 + \alpha}}{\omega_1 \omega_2 \omega_3} \frac{\pi \left( \omega_{\text{tot}}^2 + 4 m_a^2 \right)}{8} \int_{0}^{t_*} \rd t \rho(t) \int_{0}^{t_*} \rd t' \rho(t') \left[ t + t' + | t - t' | \right] \, ,
\eea
where we added the Lagrange multiplier $\mu$ for the normalization of $\rho(t)$.
In order to have a consistent saddle point $\alpha$ acquires the value $1/2$, and thus $F_{\mathbb{S}^5_\omega} \propto N^{5/2}$.
Setting to zero the variation of \eqref{S^5:free_energy:functional} with respect to $\rho(t)$ we find the following saddle point configuration
\bea
 & \rho(t) = \frac{2 | t |}{t_*} \, , \qquad t_* = \frac{1}{\sqrt{2} \sqrt{8 - N_f}} \left( \omega_{\text{tot}}^2 + 4 m_a^2 \right)^{1/2} \, , \\
 & \mu = - \frac{\pi}{3 \sqrt{2} \omega_1 \omega_2 \omega_3} \frac{N^{5/2}}{\sqrt{8 - N_f}} \left( \omega_{\text{tot}}^2 + 4 m_a^2 \right)^{3/2} \, .
\eea
Plugging this back into \eqref{S^5:free_energy:functional} we obtain the squashed $\mathbb{S}^5$ free energy of the $\USp(2N)$ theory, that reads (\cf\,\cite[(3.38)]{Chang:2017mxc})%
\footnote{The first equality arises from a virial theorem for the free energy \eqref{S^5:free_energy:functional}.}
\be
 \label{S^5:free_energy:on-shell}
 F_{\mathbb{S}^5_\omega} ( m_a | \vec{\omega}) = \frac{2}{5} \mu = - \frac{\pi \sqrt{2}}{15 \omega_1 \omega_2 \omega_3} \frac{N^{5/2}}{\sqrt{8 - N_f}} \left( \omega_{\text{tot}}^2 + 4 m_a^2 \right)^{3/2} \, .
\ee
Introducing the redundant but \emph{democratic} parameterization
\be \label{demo}
 \Delta_1 = 1 + \frac{2 \ii }{\omega_{\text{tot}}} m_a \, , \qquad \Delta_2 = 1 - \frac{2 \ii }{\omega_{\text{tot}}} m_a \, ,
\ee
\eqref{S^5:free_energy:on-shell} can be rewritten as
\be
 \label{S^5:Delta}
 F_{\mathbb{S}^5_\omega} ( \Delta_i | \vec{\omega})  = - \frac{\sqrt{2} \pi}{15} \frac{\omega_{\text{tot}}^3}{\omega_1 \omega_2 \omega_3} \frac{N^{5/2}}{\sqrt{8 - N_f}} \left( \Delta_1 \Delta_2 \right)^{3/2} \, , \qquad \Delta_1 + \Delta_2 = 2 \, .
\ee
Finally, setting $\Delta_{1,2} = 1$ and $\omega_{1,2,3} = 1$, we find the round $\mathbb{S}^5$ free energy \cite{Jafferis:2012iv}
\be
 \label{round:S^5:free_energy}
 F_{\mathbb{S}^5} = - \frac{9 \sqrt{2} \pi}{5} \frac{N^{5/2}}{\sqrt{8 - N_f}} \, .
\ee

\section[Four-dimensional solutions from the \texorpdfstring{$stu$}{stu} model]{Four-dimensional solutions from the $stu$ model}

We treat here the four-dimensional gravitational backgrounds used to compute the holographic supersymmetric R{\'e}nyi entropy. This section is organized as follows: before delving into the more intricate matter coupled solutions, we start by reviewing the simple case of the minimal supergravity BPS hyperbolic Reissner-Nordstr\"om and its SRE computation as done in \cite{Nishioka:2014mwa,Huang:2014gca}. After this, in \ref{stu4D} we first recall the basic features of four-dimensional abelian Fayet-Iliopoulos (FI) gauged supergravity and present the hyperbolic matter coupled black hole solutions which first appeared in \cite{Cvetic:1999xp}, leaving the details of the supergravity formalism and the BPS equations to appendix \ref{AppA}.  In \ref{renyi4d}, we compute the renormalized on-shell action and compare the result with the field theory computation in subsection \ref{holo3dmatching}, making contact with the minimal case as well. The complete procedure of holographic renormalization is spelled out in appendix \ref{AppB}.

\subsection[Warm up: BPS hyperbolic Reissner-Nordstrom]{Warm up: BPS hyperbolic Reissner-Nordstr\"om \label{warmup}} 
The computation of the SRE for hyperbolic solutions of $\mathcal{N}=2$ minimal gauged supergravity was treated in \cite{Nishioka:2014mwa,Huang:2014gca}. The gravity configurations are solutions to the equations of motion of the bosonic action
\be
S = \int \rd^4x \sqrt{g} \left(R - \frac14 F_{\mu\nu} F^{\mu\nu} -\frac{6}{l_{\text{AdS}}} \right) ,
\ee
and read 
\be \label{hyp_RN}
\rd s^2 = - \left(\frac{r^2}{l_{\text{AdS}}^2} -1- \frac{2M}{r} +\frac{Q^2}{r^2} \right) \rd t^2 + \frac{\rd r^2}{ \left(\frac{r^2}{l_{\text{AdS}}^2} -1- \frac{2M}{r} +\frac{Q^2}{r^2}\right)} + r^2 (\rd \theta^2 + \sinh^2(\theta) \rd \phi^2)  \, ,
\ee
with gauge field $A_t = \frac{Q}{r} \rd t+ c \? \rd t$. $c$ is a gauge term to be fixed later, in such a way that the gauge field is zero at the horizon $r_+$, where $g_{tt}$ vanishes, $g_{tt}(r_+)=0$. In order for the solution to preserve  $1/2$ of the supersymmetries, the relation $Q = \ii M $ should hold. In other words, the charges of the solution should be purely imaginary. As we elaborate later on, this is not a problem because our aim is to study an analytically continued solution in Euclidean signature, obtained by $t \rightarrow - \ii \tau$, where the metric nevertheless remains real. With a slight abuse of terminology, consistent with the literature, we will continue referring to these solutions as ``topological" or hyperbolic black holes. We set for simplicity $l_{\text{AdS}} = 1$.

First of all, imposing the BPS relation $M =- \ii Q$ and the fact that $g_{tt} (r_+) =0$ we have that
\be \label{Qis}
Q = \ii r_+ (1\pm r_+)\,.
\ee
The Wick rotated solution is characterized by a temperature $T$, found as the inverse periodicity of the $\tau$ coordinate, once we impose that the metric caps off smoothly at $r_+$. Indeed, for $r \rightarrow r_+$ the metric, upon changing coordinates to $R = \sqrt{ \frac{2(r-r_+)}{2r_+-1}} $, approaches
\be
\rd s^2 =   \rd R^2 +  R^2 \rd \tau^2 (2r_+-1) + r_+^2 (\rd \theta^2 + \sinh^2(\theta) \rd \phi^2) \, .
\ee
Therefore, the periodicity of the $\tau $ coordinate should be $\beta \equiv \Delta \tau = \frac{2\pi}{2r_+-1}$. The temperature\footnote{Once again this we denote this as "temperature of the black hole" but indeed we stress that its meaning comes from the Euclidean solution.} is the inverse of this period:
\be \label{T4min}
T = \frac{2r_+-1}{2\pi} \, .
\ee
In order for the gauge field not to be singular at the horizon
\be
A(r_+) = \frac{Q}{r_+} \rd t+ c \? \rd t =0 \, ,
\ee
we set $c= - \frac{Q}{r_+}$. We define the chemical potential $\phi$ as the asymptotic value of the gauge field, therefore $\phi \equiv \lim_{r \rightarrow \infty} A_{t} = c$. 

To find the SRE, one identifies $T$ with $T_0/n$, where $T_0$ is the temperature of the neutral black hole and $n$ is the replica parameter. In this way,
\be \label{Treplica}
T = \frac{1}{2\pi n} \, .
\ee
Combining \eqref{T4min} and \eqref{Treplica}, we can extract the value of $r_+$ as a function of the replica parameter $n$:
\be \label{rns}
r_+ = \frac{n \mp1}{2n} \, .
\ee
We choose the lower branch since, for $n=1$, $r_+$ should go to unity. Similar reasoning makes us choose the lower sign in \eqref{Qis}. The expression for the free energy found in \cite{Nishioka:2014mwa,Huang:2014gca}  reads
\be \label{onsh_min}
I = \frac{\Vol(\bH^2) \beta}{8 \pi G_4} \left(-r_+^3 + \ii Q - \frac{Q^2}{r_+} \right) \, ,
\ee
which, upon using  \eqref{Qis}, \eqref{rns} becomes 
\be \label{onsh}
 I = \frac{\Vol(\bH^2) \beta}{8 \pi G_4}  \frac{(n+1)^2}{n^2}
 = \frac{ \pi }{8 G_4} \frac{(n+1)^2}{n} \, .
\ee
This matches the branched sphere partition function on the field theory side \cite{Nishioka:2014mwa,Huang:2014gca}, upon setting $\Delta_I = 1/2$, $I=1,\ldots,4$, in \eqref{br_sph} and using the standard AdS$_4$/CFT$_{3}$ relation $\frac{1}{G_4} = \frac{2 \sqrt2}{ 3} N^{3/2}$ and the regularized volume $\Vol(\bH^2) =-2\pi$ \cite{Nishioka:2014mwa}. 

Finally, we notice that the chemical potential takes the form
\be
\phi = - \frac{Q}{r_+}= - \ii (1- r_+) = - \ii \frac{n-1}{2n} \, ,
\ee
matching the value of the R-symmetry background field \eqref{Rsymm}. We record this expression as it will be useful later on in the computation of the SRE in the matter coupled case.

\subsection[Hyperbolic black hole solutions of the $stu$ model]{Hyperbolic black hole solutions of the $stu$ model}
\label{stu4D}

The AdS$_4$ black holes with hyperbolic horizon we are after are solutions to abelian FI gauged supergravity in four spacetime dimensions. $\U(1)$ FI gauged supergravity arises as a truncation to the Cartan subalgebra, $\U(1)^4$, of $\mathcal{N} = 8$ gauged supergravity. The model thus obtained, called the $stu$ model, corresponds to the prepotential 
\begin{equation}\label{prep}
F ( X ) = - 2 \ii \sqrt{X^0 X^1 X^2 X^3 } \, ,
\end{equation}
in the standard notation of $\mathcal{N} =2$ supergravity. We will deal with a purely electric solution that has a hyperbolic horizon, supported by purely real scalars. In the BPS limit, the solution correspond to a $1/2$ BPS black hole, preserving 4 out of the original 8 supercharges.

Spherical black holes of this model were constructed in  \cite{Duff:1999gh, Sabra:1999ux}, and later elaborated upon in \cite{Toldo:2012ec}. The hyperbolic solution, along with its uplift to eleven dimensions, first appeared in \cite{Cvetic:1999xp}. It is a static black hole characterized by the following metric
 \begin{equation}\label{sol}
\rd s^2= -\frac{U(r)}{4} \rd t^2 +\frac{\rd r^2}{U(r)}+ h^2(r) ( \rd \theta^2 + \sinh^2(\theta) \rd \phi^2) \,,
\end{equation}
with
\begin{equation} \label{warpp3}
U(r)=\frac{1}{\sqrt{\mathcal{H}}} f(r) \,, \qquad f(r) = -1- \frac{\mu}{r}+4 g^2 r^2 \mathcal{H} \,,
\qquad h^2(r)=\sqrt{\mathcal{H}} r^2 \,.
\end{equation}
and
\be
\mathcal{H} = H_1 H_2 H_3 H_4 \, , \qquad H_{I} = 1 + \frac{b_{I}}{r} \, , \qquad I =1, \ldots, 4 \, .
\ee
We set $g=1$ from now on, and notice that we have rescaled time to match the asymptotic geometry \eqref{hypn}. The non-vanishing components of the vector fields supporting the configurations are
\begin{equation}\label{gf}
A^{I} = \frac12 \left(1-\frac{1}{H_I}  \right)  \frac{q_I}{b_I} \, \rd t + c^I \rd t\, ,
\end{equation}
where we have included four constant parameters $c^I$ (to be determined later) which are required so that the gauge fields are non-singular at the horizon. 

The equations of motion are satisfied if the parameters satisfy the following relation:
\be
b_I =  \mu \sin^2(\zeta_I) \, , \qquad q_I= \mu \sin(\zeta_I) \cos(\zeta_I) \, .
\ee
Uppercase indices $I,J$ run from 1 to 4, while lowercase ones $i,j$ run from 1 to 3.
The magnetic charges are set to zero, hence this is a purely electric configuration. The scalar fields $z^i$ are \emph{real} and parameterized by the holomorphic sections $X^{i}$, $z^i = X^i / X^0$. They assume the form \cite{Duff:1999gh}
\be \label{zis}
z^1 = \frac{H_1  H_2}{H_3 H_4} \, , \qquad z^2 = \frac{H_1  H_3}{H_2 H_4} \, , \qquad z^3= \frac{H_1  H_4}{H_2 H_3} \, .
\ee
The uplift of the solution to eleven-dimensional supergravity was performed in \cite{Cvetic:1999xp}, where the solution was interpreted as the decoupling limit of spinning M2-branes. The BPS branch, which provides the solutions of interest here, is obtained by setting $\mu=0$ and by taking%
\be
q_I= \ii b_I \, .
\ee 
This configuration solves the BPS equations, as shown in appendix \ref{4D:BPS:proof}. Notice that the electric charge assumes a purely imaginary value, as it did in the minimal case studied in \cite{Nishioka:2014mwa,Huang:2014gca}. This is not a problem, as our aim is to study an analytically continued solution preserving supersymmetry. For this purpose, it is legitimate to take some parameters to be genuinely complex, since the Killing spinor equation, being analytic in the supergravity fields, will still admit a solution in the complexified background. Nevertheless, the Euclideanized metric in this case will remain purely real. It would be desirable to find a suitable solution directly in Euclidean supergravity coupled to matter multiplets, however in the following we will content ourselves with (a Wick-rotated version of) the Lorentzian solutions at hand.

The hyperbolic Reissner-Nordstr\"om solution discussed in the previous subsection is recovered from our setup upon taking the scalars to be constant
\be
H_1=H_2=H_3=H_4 =H \, , \qquad z^i =1 \, ,\quad  i = 1,2,3 \, ,
\ee
taking all the gauge fields equal, and redefining the $stu$ fields $A^I$ (see \cite[(3.15)]{Cvetic:1999xp}) as $A^I = A / 2$. By doing so, the number of independent electric charges reduces to one, that of the graviphoton $A$.

\subsection[Holographic supersymmetric Renyi entropy]{Holographic supersymmetric R{\'e}nyi entropy \label{renyi4d}}

From the $stu$ black hole at our disposal, we can compute the temperature (see footnote 14)
\be
T = \frac{1}{4\pi} \frac{\rd U}{\rd r} \bigg|_{r_+} \, ,
\ee
which turns out to be 
\be \label{temp}
T = \frac{ \left(r_+^3 (b_3+b_4+2 r_+)-b_1 \left(b_2 b_3 (2 b_4+r_+)+b_2 b_4 r_++b_3 b_4 r_+-r_+^3\right)+b_2 \left(r_+^3-b_3 b_4 r_+\right)\right)}{2 \pi  r_+ \sqrt{b_1+r_+} \sqrt{b_2+r_+} \sqrt{b_3+r_+} \sqrt{b_4+r_+}} \, .
\ee
Here, $r_+$ is the location of the horizon, obtained by requiring $U(r_+) =0$. We leave the quantity $r_+$ implicit for the moment:
trying to solve for $r_+$ from the vanishing of the warp factor yields a quartic equation whose explicit expression is quite cumbersome to manipulate.

Consider the uncharged black hole $q_1=q_2=q_3=q_4=0$. In this case, the requirement $U(r_+) =0$ gives  $4 r_+^2  - 1=0$, hence $r_+$ takes the simple form
\be
 r_+ = \frac{1}{2} \, .
\ee
Denoting by $T_0$ the temperature of the uncharged black hole, we have
\be
T_0 = \frac{1}{2\pi} \, ,
\ee 
which will be useful later when defining the supersymmetric R{\'e}nyi entropy.

In order for the gauge field to be non-singular at the horizon, we require $A^I(r_+) =0$. Given the expression \eqref{gf},  this leads to
\be
c^I = - \frac{\ii}{2} \left(1-\frac{1}{H_I (r_+)}  \right) , \qquad I = 1, \ldots, 4 \, .
\ee
The chemical potentials $\phi_I$ are defined as the asymptotic values of the gauge fields.
They assume the form (we do not distinguish here between upper and lower indices on the chemical potentials)
\be \label{potential}
\phi_I = c^I 
= -\frac{\ii}{2} \frac{b_I}{ b_I + r_+} \, , \qquad I = 1, \ldots, 4 \, .
\ee

By inserting \eqref{potential} into \eqref{temp}, we can express the temperature as a function of the chemical potentials in the following way:
\be
T= \frac{- \ii (\phi_1+\phi_2+\phi_3+\phi_4) + 1}{ \pi ( \sqrt{1-2 \ii \phi_1} \sqrt{1-2 \ii \phi_2} \sqrt{1- 2 \ii \phi_3} \sqrt{1- 2 \ii \phi_4})} \? r_+ \, ,
\ee
where we have once again left $r_+$ implicit. We also point out that the quantities $\phi_I$ are \emph{imaginary}, therefore $T$ is \emph{real}, as it should be.
At this point, we can define
\be \label{renyi}
T = \frac{T_0}{n} = \frac{1}{2 \pi n} \, .
\ee
Solving this equation for $r_+$ we obtain
\be \label{rplus4}
r_+=\frac{1}{2 n} \frac{\sqrt{1- 2\ii \phi_1} \sqrt{1- 2 \ii \phi_2} \sqrt{1- 2\ii \phi_3} \sqrt{1-2 \ii \phi_4}}{1-\ii (\phi_1+\phi_2+\phi_3+\phi_4) } \, .
\ee
Additionally, we know that the quantity $r_+$ must satisfy the relation $U(r_+) =0$.
Inserting the definitions \eqref{potential} into $U(r_+) =0$ yields the condition
\be
1+n^2 (\phi_1 + \phi_2 + \phi_3 +\phi_4+ \ii)^2 =0 \, ,
\ee
which is solved by
\be \label{constr_chempot}
\phi_1 + \phi_2 + \phi_3 +\phi_4 = \frac{ \ii ( 1\pm n )}{n} \, .
\ee
We choose the lower sign since for $n=1$ we should have zero chemical potential. As we will see in a moment, the choice of the upper branch translates in the dual field theory to a constraint on the value of the R-symmetry background field. To recapitulate, at this point we have obtained the expression \eqref{rplus4} for $r_+$ in terms of the chemical potentials and the R{\'e}nyi parameter $n$, supplemented by the constraint \eqref{constr_chempot}.

The renormalized on-shell action is computed by adapting the procedure of \cite{Batrachenko:2004fd} to the case of hyperbolic horizons.
The computation, reported in appendix \ref{AppB}, is tedious and not particularly illuminating. In the end, the thermodynamical potential reads
\begin{eqnarray}\label{Omega}
I &=&\beta \Omega  = I_{\text{reg}}+ E_{\text{ct}}+E_{\text{fin}}= \frac{\beta \, \Vol(\bH^2)}{8 \pi \textbf{c} G_4}\left(-\frac{\mu}{2}+r_+\right) \, ,
\end{eqnarray}
where $I_{\text{reg}}$ is the regularized on-shell action, $I_{\text{ct}}= E_{\text{ct}}$, $\Vol(\bH^2)$ is the (regularized) volume of $\bH^2$, and $\beta = 1/T$ is the period of the Euclidean time direction.
For the BPS case $\mu =0$, we have 
\bea\label{Omega}
I = \frac{\beta \? \Vol(\bH^2) \, r_+}{8 \pi \textbf{c} G_4} =\frac{ 2 \pi  \Vol(\bH^2)}{8 \pi \textbf{c} G_4} \left(\frac{ \ii \sqrt{1-2 \ii \phi_1} \sqrt{1-2 \ii \phi_2} \sqrt{1-2 \ii \phi_3} \sqrt{1-2 \ii \phi_4}}{  2  ( \ii + \phi_1+\phi_2+\phi_3+\phi_4)}
 \right) .
\eea
This expression is useful when comparing with the field theory result $Z_n$ \eqref{br_sph}. The free energy of the black hole is given by \be
I = - \log Z(\phi_I,T) \, .
\ee
The state variables are computed according to
\be \label{rel1}
E = \left( \frac{\partial I}{ \partial \beta}\right)_{\phi} - \frac{\phi_I}{\beta} \left( \frac{\partial I}{\partial \phi_I}\right)_{\beta} \, ,
\qquad 
S_{\text{BH}} = \beta \left(\frac{\partial I}{ \partial \beta} \right)_{\phi} - I \, ,
 \qquad 
\mathcal{Q}_I = -\frac{1}{\beta} \left(\frac{\partial I}{ \partial \phi_I} \right)_{\beta} \, .
\ee
The renormalized on-shell action, \eqref{Omega}, is computed in the grand canonical ensemble. In this ensemble, the Gibbs potential $W$ is given by (see appendix \ref{AppB})
\be
W = \frac{I}{\beta} = E - T S_{\text{BH}} - \phi^I Q_I + \Lambda \left( \phi_1+\phi_2+\phi_3+\phi_4 - \ii \frac{(1-n)}{n} \right) \, ,
\ee
where $Q_I$ are the electric charges of the black hole, and we inserted the Lagrange multiplier $\Lambda$ which enforces the constraint \eqref{constr_chempot} among the chemical potentials.

\subsection{Holographic matching \label{holo3dmatching}}

In this section, we perform the holographic matching. The asymptotic value of the four-dimensional bulk gauge fields is related to the dual field theory flavor symmetry connection, defined in section \ref{def}, as 
\be
 A_{\text{bulk}}^I (r \to \infty)= \phi_I \rd t = \big( A^{\text{flavor},I}({\mathbb{S}^3_n})+A^{(R)}_{\text{bulk}} \big)\rd \tau \, ,
\ee
where we have used $t = -\ii \tau$. To preserve supersymmetry, the background R-symmetry gauge field must have the form \eqref{Rsymm}. The background R-symmetry gauge field is identified with the chemical potential related to the R-symmetry gauge field in supergravity, which is the diagonal combination\footnote{Note that the factor of $1/2$ between \eqref{Rcomb} and \eqref{Rsymm} is due to the fact that the gauge fields in the $stu$ model are defined with a factor of 
$1/2$ with respect to the graviphoton in minimal supergravity \cite{Cvetic:1999xp}.}
\be \label{Rcomb}
A^{(R)}_{\text{bulk}} (r \rightarrow \infty) = \frac14 \left(\phi_1 + \phi_2 +\phi_3 +\phi_4 \right) \rd t = \ii \frac{1-n}{4n} \rd t = \frac{1-n}{4n} \rd \tau \, ,
\ee
that appears in the supercovariant derivative of the spinor parameter in the susy variations \eqref{susyvar4D}. Notice that $A^{(R)}_{\text{bulk}} = \frac12 A^{(R)}$. As a simple consistency check, \eqref{Rcomb} is precisely the relation \eqref{constr_chempot} obtained previously.%

We are now ready to make contact with the field theory.  The bulk fields correspond to the holonomies, shifted by the amount $(1 - n) / (4n)$ due to the R-symmetry connection. In other words, we use the mapping \eqref{map3d}  between the holonomies $A^{\text{flavor},I}$ and the parameters $\Delta_I$, supplemented by the shift due to the R-symmetry:

\be
A^I = A^{\text{flavor},I}+ A^{(R)}_{\text{bulk}} =  \left( \Delta_I -\frac12 \right) \left( \frac{n+1}{2n}\right) + \frac{1-n}{4n} =  \left(\frac{(1+n)\Delta_I}{2n} -\frac12 \right) .
\ee
Thus, we have  
\be \label{mapphi3}
\phi_I = \ii \left( \frac{(1+n)\Delta_I}{2n} -\frac12 \right) ,  \qquad I =1,...,4 \, .
\ee
Taking the sum of the LHS and the RHS we obtain the constraint
\be
\frac{n-1}{n} = 2 - \frac{n+1}{2n} \sum_I \Delta_I \qquad \Rightarrow \qquad \sum_I \Delta_I=2 \, ,
\ee 
which reproduces the usual constraint on the parameters $\Delta_I$. We use the standard relation
\be
 \label{AdS4:CFT3:dict}
 \frac{l_{\text{AdS}}^2}{G_4} = \frac{2 \sqrt2}{3} N^{3/2} \, ,
\ee
where we have taken into account $l^2_{\text{AdS}} =1/4$ from \eqref{warpp3}. Inserting \eqref{mapphi3} into \eqref{Omega}, with $\textbf{c}=2$, the expression of the free energy becomes
\be
 \label{s3}
I=-  \frac{ \sqrt2 \pi N^{3/2}}{3}  \frac{(n+1)^2}{n} \sqrt{\Delta_1 \Delta_2 \Delta_3 \Delta_4} =  - \log Z_{\mathbb{S}_n^3} \, , \qquad \sum_{I=1}^4 \Delta_I =2 \, ,
\ee
exactly matching the field theory computation \eqref{br_sph} upon identifying $b \equiv 1/\sqrt{n}$, see \eqref{map3d}. Note that we have defined the regularized volume as $\Vol(\bH^2) = -2\pi$ as in \cite{Nishioka:2014mwa}. One easily sees that at the conformal point, $\Delta_I=1/2$, which corresponds to the minimal supergravity case, the on-shell action reduces as expected to
the one found in \cite{Nishioka:2014mwa,Huang:2014gca}.

We are now going to compute the supersymmetric R{\'e}nyi entropy. First, notice that the partition function on the field theory side, see \eqref{br_sph}, satisfies
\be
 \log Z_{\mathbb{S}_n^3} = \frac{(n+1)^2}{4n} \log Z_{\mathbb{S}^3} \, .
\ee
The supersymmetric R{\'e}nyi entropy is defined as \eqref{def:SRE} \be
S_n^{\text{SRE}} = \frac{n \log Z_{\mathbb{S}^3} - \log Z_{\mathbb{S}_n^3}}{n-1} \, .
\ee
Therefore, we have
\be \label{renyientropy}
S_n = \frac{3n+1}{4n} S_1 \, , \qquad S_1 = \log Z_{\mathbb{S}^3} \, ,
\ee
as expected. 

\section[Six-dimensional hyperbolic solutions]{Six-dimensional hyperbolic solutions}

We introduce here the six-dimensional hyperbolic solutions necessary for the holographic computation of the supersymmetric R{\'e}nyi entropy. We first give some details regarding six-dimensional Romans $F(4)$ gauged supergravity coupled to one vector multiplet. We then present the hyperbolic black hole solutions coupled to matter, which have not previously appeared in the literature. In \ref{SRE6dROM}, we compute the holographic R{\'e}nyi entropy, using the result of appendix \ref{AppB}, and show the matching with the field theory computation in section \ref{sec:logZ:5D}.

\subsection[Romans F(4) gauged supergravity coupled to matter]{Romans F(4) gauged supergravity coupled to matter}

In what follows, we will consider the six-dimensional  F(4) gauged supergravity coupled to one vector multiplet. Relevant references for this theory are \cite{Andrianopoli:2001rs,DAuria:2000xty}. While the massive type IIA supergravity origin of this theory as a truncation of the supersymmetric warped AdS$_6 \times \mathbb{S}^4$ solution has not been established, there is evidence for it based on previous holographic matchings, see for instance \cite{Gutperle:2017nwo,Hosseini:2018usu}. Taking the pragmatic approach of these latter papers, we work out supersymmetric solutions and proceed with the comparison of our result with its field theory counterpart. The  five-dimensional SCFT dual  to  the warped   AdS$_6 \times \mathbb{S}^4$ background is the one described in section \ref{sec:logZ:5D}. Solutions relevant for the supersymmetric R{\'e}nyi entropy computation in the minimal theory (no vector multiplets) \cite{Romans:1985tw} were studied in \cite{Hama:2014iea,Alday:2014fsa}. The non-minimal case is characterized by the presence of an additional flavor symmetry.

The bosonic fields of the six-dimensional Romans supergravity theory \cite{Romans:1985tw} consist of the metric $g_{\mu \nu}$, a scalar field $X$, a two-form potential $B_{\mu \nu}$, a one-form potential $A$,
and an $\SU(2)$ gauge field $A^j$ with $j= 1,2,3$. In addition, there are fermionic fields comprising a pair of gravitini $\psi_{\mu}^A$, $A=1,2$ and one spin 1/2 fermion $\chi^A$.
The vector multiplets consist of one gauge field $A_{\mu}$, four scalar fields $\phi_{\alpha}$, with $\alpha=0,1,2,3$, and one gaugino $\lambda_A$. The scalar fields parameterize the coset space $\frac{\SO(4,1)}{\SO(4)}$. For additional details on the model, we refer the reader to \cite{Gutperle:2017nwo,Hosseini:2018usu}.

In finding the solution, we may take the Romans supergravity solution as example.
In this solution, only one of the components of the $\SU(2)$ gauge field, which we take to be $A^3$ \cite{Alday:2014fsa}, is nonzero. This gauge field is purely electric, meaning that the only nonzero component of the field strength is $F_{rt}$.
This allows us to set the two-form potential $B_{\mu \nu}$ to zero, as there is no source for it.%
\footnote{This is in contrast to the six-dimensional solutions of \cite{Hosseini:2018usu} of the form AdS$_2 \times \Sigma_{g_1} \times \Sigma_{g_2}$, which realizes the partial topological twist on $\Sigma_{g_1} \times \Sigma_{g_2}$. In that case, there is magnetic flux on $\Sigma_{g_1}$ and $ \Sigma_{g_2}$. This creates a source for the $H_{\mu \nu}$ field, which needs to be canceled by a nonzero value of $B$, in order to have a solution with $H=0$.}
In our setup with an additional vector multiplet, we will still require the $B$ field to vanish.
Moreover, as in \cite{Hosseini:2018usu}, we require the scalar fields in the vector multiplet $\phi_{\alpha}$ to be neutral under $A^3$. This restricts the nonzero components to $\phi_0$ and $\phi_3$.
We are further able to find a solution with only $\phi_3$ turned on, namely $\phi_0=0$. Thus, we are left with the bosonic content: the metric, two gauge fields, the dilaton $X$, and the scalar field $\phi_0$.

\subsection[Supersymmetric hyperbolic black holes]{Six-dimensional supersymmetric hyperbolic black holes}\label{6dF4}

For the non-minimal case, we adapt the solutions of \cite[sect.\,3.2]{Chow:2011fh} to the $\mathbb{H}^4$ horizon topology. 
The solution is a static black hole characterized by the following metric
\begin{equation}
\rd s^2= -U(r) \rd t^2 +\frac{\rd r^2}{V(r)}+ h(r) \rd s_{\mathbb{H}^4}^2 \, ,
\end{equation}
with $\rd s^2_{\mathbb{H}^4}$ the area element of four-dimensional hyperbolic space
\be
\rd s^2_{\mathbb{H}^4} = \rd \chi^2 +\sinh(\chi)^2 \left( \rd \theta^2 +\sin^2(\theta) \rd \psi^2 + \sin^2 (\theta) \sin^2(\chi) \rd \phi^2 \right) ,
\ee
and
\begin{equation} \label{6D_warp}
U(r)=\frac92 \frac{f(r)}{\mathcal{H}^{3/4}}\,, \qquad V(r)= \frac{f(r)}{\mathcal{H}^{1/4}} 
\qquad h(r)=\mathcal{H}^{1/4} r^2 \, ,
\end{equation}
with%
\footnote{As in \cite{Alday:2014fsa}, we have conveniently rescaled the time direction by a factor of $3/\sqrt2$ with respect to \cite{Cvetic:1999un}.}
\be
f(r) = -1- \frac{\mu}{r}+ \frac29 r^2 \mathcal{H}  \qquad \mathcal{H} = H_1 H_2 \, , \qquad H_I = 1 + \frac{b_I}{r^3} \, .
\ee
Here, $I=1,2$. The vector fields supporting the configuration read
\begin{equation}\label{AI6}
A^{I}_t = \frac{3}{2} \left(1-\frac{1}{H_I}  \right) \frac{q_I}{b_I}\, - c^I \rd t , \qquad I=1,2 \, ,
\end{equation}
with parameters
\be
b_I = \mu \sin^2(\xi_I) \, , \qquad q_I = \mu \sin(\xi_I) \cos(\xi_I) \, ,
\ee
and the scalars, in the notation of \cite{Chow:2011fh} are given by
\be
X_1 = H_1^{-5/8} H_2^{3/8}\,, \qquad  X_2 = H_1^{3/8} H_2^{-5/8} \,.
\ee
The configuration with spherical slicing first appeared in \cite{Chow:2011fh}, and the solution presented here is its generalization to hyperbolic slicing.
However, the origin of the original configuration as a solution of a supergravity theory was unclear. It is easy to verify that the configuration is a solution to the equations of motion of F(4) gauged supergravity coupled to one vector multiplet, which are reported in \cite{Suh:2018szn}. One first truncates the theory to the $\U(1) \times \U(1)$ sector, as was done in \cite{Karndumri:2015eta}, obtaining the Lagrangian \cite[(3.2)]{Suh:2018szn}. One can then see that the field $\varphi_1$ can be consistently set to zero.
Moreover, since all the field strengths are electric, there is no source term for the field $B_{\mu\nu}$, hence the latter can be set to zero as well.
The remaining fields in our solutions can be mapped to those in \cite{Hosseini:2018usu,Suh:2018szn} via%
\footnote{The field we call $\phi_3$ and $F_{i1}$ coincides respectively with $\phi_2$ and $F_6$ of \cite{Suh:2018szn}.}
\bea
 & F_1 = \rd A_1= F_3 -F_{i1} \, , \qquad && F_2 = \rd A_2 =F_3 + F_{i1} \, , \\
 & X_1 =  e^{\sigma-\phi_3} \, , && X_2 = e^{\sigma + \phi_3} \, .
\eea
With this mapping, and once we impose the truncations described above, one can show that the equations of motion are solved. The gauging parameters $g,m$ are set to $g=3m$ and $m=1/(3 \sqrt2)$, justifying the factor $2/9$ in the warp factor $f(r)$ in \eqref{6D_warp}.

The BPS branch is obtained, as usual, by setting $\mu =0$ and $q_I = \ii b_I$. The solution is $1/2$ BPS, and its Killing spinor is explicitly constructed in \ref{6D:BPS:proof}. These solutions, once a Wick rotation to Euclidean spacetime is performed and setting $b_1=b_2$, reduce to those considered in \cite{Hama:2014iea,Alday:2014fsa}. 

\subsection[Supersymmetric Renyi entropy]{Supersymmetric R{\'e}nyi entropy \label{SRE6dROM}}

As in the previous case, we start the procedure by computing the period of the Euclidean time circle, namely the temperature of the hyperbolically sliced black hole. Given the expression for the warp factor \eqref{6D_warp}, we have 
\be
T=-\frac{\left(4 b_1 b_2+ b_1 r_+^3+ b_2 r_+^3-2 r_+^6\right)}{6 \sqrt2 \pi  r_+^2 \sqrt{ b_1+r_+^3} \sqrt{ b_2+r_+^3}} \, .
\ee
Once we impose that the gauge field vanishes at the black hole horizon, we introduce the chemical potentials $\phi_I$, $I=1,2,$ as the asymptotic value of the gauge fields \eqref{AI6}. We obtain
\be
\phi_I = - \frac32 \frac{q_I}{b_I + r_+^3}  = -\frac32 \frac{\ii b_I}{b_I + r_+^3} \, , \qquad I = 1,2 \, ,
\ee
where in the second equality we have used the BPS relation $q_I = \ii b_I$.
The temperature can then be rewritten as
\be
 T=\frac{1}{\sqrt2 \pi} \frac{1- \ii (\phi_1+  \phi_2) }{\sqrt{3-2 \ii \phi_1} \sqrt{3-2 \ii \phi_2}} \? r_+ \, .
\ee 
By equating $T = T_0/n= 1 / (2 \pi n )$, we obtain an expression for $r_+$ in terms of the chemical potentials and the R{\'e}nyi parameter $n$:
\be \label{rplus6}
r_+ =  \frac{ \sqrt{3- 2 \ii \phi_1} \sqrt{3-2 \ii \phi_2}}{  \sqrt2 n(1 - \ii ( \phi_1+ \phi_2))} \, ,
\ee
taking into account once more that these quantities are related via
\be
 \label{constr_chem6}
 \phi_1 + \phi_2 =  \frac{ \ii (1 \pm n)}{n} \, .
\ee
As explained in the previous section, we choose the lower sign so that the configuration reduces to a neutral black hole for $n=1$. 

The renormalized on-shell action can be computed easily (see appendix \ref{AppB}) by imposing supersymmetry. Using $\textbf{c} = \sqrt2/3$, we obtain
\be \label{onsh6}
I = \frac{\beta \Vol(\mathbb{H}^4)}{8 \pi \textbf{c} G_6}  \left( - r_+^3 - \frac{\mu}{2} \right)  = - \frac{3  n}{4 \sqrt2   G_6} \Vol(\mathbb{H}^4) r_+^3 \, .
\ee
This is consistent with the result of \cite{Alday:2014fsa}, which is valid in the absence of vector multiplets.
\eqref{onsh6} combined with the previous expression, \eqref{rplus6} for $r_+$ yields
\be \label{fin_grav_6}
I
=  \frac{ \pi^2 n}{\sqrt2 G_6} \left(  \frac{ \sqrt{3- 2 \ii \phi_1} \sqrt{3-2 \ii \phi_2}}{  \sqrt2 n(1 - \ii ( \phi_1+ \phi_2))} \right)^3 ,
\ee
supplemented by the constraint \eqref{constr_chem6} between the chemical potentials. We have also used the normalized volume
$
\Vol(\mathbb{H}^4) = 4 \pi^2 /3
$
\cite{Alday:2014fsa}.

\subsection[Holographic matching]{Holographic matching}

We recall the expression that relates the asymptotic value of the bulk gauge field to the corresponding dual quantities:
\be
 A_{\text{bulk}}^I (r \to \infty) = \phi_I \rd t = \big( A^I({\mathbb{S}^5_n}) +A_{\text{bulk}}^{(R)} \big) \rd \tau\, .
\ee
Recall that, on the field theory side, the R-symmetry background gauge field has the expression \eqref{Rsymm}. The corresponding chemical potential in the supergravity notation reads
\be
A^{(R)}_{\text{bulk}} = \frac{\phi_1 + \phi_2}{2} \rd t = \ii \frac{1-n}{2n}  \rd t = \frac{ 1-n}{2n} \rd \tau\, .
\ee
We are ready now to make contact with the field theory chemical potentials. Indeed,
the bulk fields correspond to \eqref{6d_mapping}, which are related to $\Delta_I$ via \eqref{demo}, shifted by the amount $(1-n) / (2n)$ due to the R-symmetry connection, resulting in
\be
A^I=  \left( \Delta_I -1 \right) \left( \frac{2n+1}{2n}\right) + \frac{1-n}{2n} = \frac32 \left(\frac{(1+2n)\Delta_I}{3n} -1 \right) .
\ee
Therefore, we have  
\be \label{mapphi}
\phi_I = \ii \frac32 \left(\frac{\Delta_I (2 n+1)}{3 n} -1\right) , \qquad I =1,2 \, .
\ee
Notice that taking the sum over the index $I$ and using \eqref{constr_chem6} we get the relation $\Delta_1 +\Delta_2 = 2$.  
Taking into account \eqref{mapphi}, noting that $l_{\text{AdS}}^2= 9/2$, and using the relation \cite{Jafferis:2012iv}
\be
\frac{l_{\text{AdS}}^4}{G_6} =  \frac{27 \sqrt2 }{  \sqrt{8-N_f}} \frac{N^{5/2}}{5 \pi} \, ,
\ee
the gravitational on-shell action in \eqref{fin_grav_6} yields exactly
\be \label{final_6D}
I
=\frac{ \sqrt2 \pi N^{5/2}}{15 \sqrt{8-N_f} } \frac{(2 n+1)^3}{n^2} (\Delta_1 \Delta_2)^{3/2} \,, \qquad \Delta_1+ \Delta_2 =2 \, .
\ee
This perfectly agrees with the prediction from the field theory \eqref{S^5:Delta}, once we set $\vec{\omega} = (1,1,1/n)$. In the absence of flavor symmetry (or masses), we obtain the result of the minimal case. 
Indeed, imposing $\Delta_1 = \Delta_2 =1$ we retrieve the result of \cite{Hama:2014iea,Alday:2014fsa}, which reads
\be \label{res_min}
I =\frac{ \sqrt2 \pi  \? (2 n+1)^3 N^{5/2}}{15 n^2 \sqrt{8-N_f}} = - \log Z_{\mathbb{S}_n^5}  \,.
\ee

One can easily work out the value of $S_n^{\text{SRE}}$ as
\be \label{sq6}
S_n^{\text{SRE}} =  \frac{n \log Z_{\mathbb{S}^5} - \log Z_{\mathbb{S}_n^5}}{n-1}=  \frac{19n^2+7n+1}{27n^2} S_1 \, , \qquad S_1 = \log Z_{\mathbb{S}^5} \, .
\ee

\section{Concluding remarks}

Following the work on magnetically charged AdS$_4$ black holes in \cite{Benini:2015eyy}, intense efforts have been put into the holographic computation
of entropy for BPS black holes with compact horizons, using localization (see \cite{Zaffaroni:2019dhb,Hosseini:2018qsx} and references within).
Some of the computations involve a rather subtle treatment of the matrix integrals which compute the relevant SCFT partition function.
For instance, progress has been made on the longstanding problem of computing the entropy of rotating BPS black holes in AdS$_5$ from the superconformal index of $\mathcal{N}=4$ SYM using such a treatment \cite{Benini:2018ywd}. Our computation is somewhat similar, the black holes in question having no magnetic flux, but does not involve the same subtleties. This may be due to the observation, made in \cite{Cabo-Bizet:2018ehj}, that the Killing spinors relevant to the computation in the bulk, and hence in the SCFT, should be anti-periodic in the Euclidean time direction. While this can be arranged for partition functions like the one used to compute the superconformal index \cite{Closset:2017zgf}, it arises naturally in the context of the SRE, \ie\,the hyperbolic index, when viewed as a Weyl transformation of the branched sphere. This fact still awaits a satisfactory physical explanation.

Regarding possible future directions, it would be interesting to incorporate magnetic charges in the black hole background, and compare the resulting free energy with the corresponding field theory computation generalized by magnetic fluxes. Moreover, one could compute the subleading $N$ corrections to the Supersymmetric R{\'e}nyi and compare with the supergravity computation, along the lines of \cite{Nian:2017hac}. Finally, it would be interesting to investigate in our setup the expansion of the SRE around $n=1$. In \cite{Closset:2012ru,Perlmutter:2013gua,Nishioka:2013haa} it was found that the first correction to the entanglement entropy is proportional to the coefficient of the stress tensor vacuum two-point function, and it would be interesting to find the interpretation of this statement in the supergravity picture. We hope to come back to these points in the future.

\section*{Acknowledgements}

We would like to thank Laura Andrianopoli, Davide Cassani, Martin Fluder, M\'ark Mezei, Ioannis Papadimitriou, Julian Sonner for discussions, Kiril Hristov, Tatsuma Nishioka and Alberto Zaffaroni for carefully reading a first draft of the manuscript. The work of SMH was supported by World Premier International Research Center Initiative (WPI Initiative), MEXT, Japan.
CT acknowledges support from the NSF Grant PHY-1125915 and the Agence Nationale de la Recherche (ANR) under the grant Black-dS-String (ANR-16-CE31-0004)
and would like to thank the Simons Center for Geometry and Physics, Stony Brook University, Kavli IPMU and Universita' di Parma for hospitality during some steps of this paper. The work of IY was financially supported by the European Union's Horizon 2020 research and innovation programme under the Marie Sklodowska-Curie grant agreement No. 754496 - FELLINI.

\appendix

\section{Explicit construction of the Killing spinor \label{AppA}}

\subsection{Four dimensions}
\label{4D:BPS:proof}

To show the resolution of the Killing spinor equations (KSE) of four-dimensional abelian FI gauged supergravity in presence of vector multiplets we follow essentially the conventions of \cite{Sabra:1999ux}, where the Killing spinor for configurations with spherical spatial section was worked out (see also \cite{Duff:1999gh}). The modification to configurations with hyperbolic horizon is an easy task that we perform in what follows.
 
The supersymmetry transformation of the gravitini and gaugini in terms of complex spinors read 
\bea
 \label{susyvar4D}
 \delta\psi_{\mu} & =\nabla_{\mu} \varepsilon + \frac{\ii}{4} T_{\rho \sigma}^- \gamma^{\rho} \gamma^{\sigma} \gamma_{\mu} \varepsilon - \frac{g}{2}  \xi_{\Lambda} L^{\Lambda} \gamma_{\mu} \varepsilon \, , \\
 \delta_{\varepsilon } \lambda^{i} & =  \ii \partial_{\mu} z^i \gamma^{\mu} \varepsilon + \ii G_{\mu \nu}^{-i} \gamma^{\mu \nu} \varepsilon +  g g^{i \bar{\jmath}} \bar{f}_{\bar{\jmath}}^{\Lambda} \xi_{\Lambda} \varepsilon \,.
\eea
The supercovariant derivative of the gravitino appearing in the supersymmetry variations is given by
\be
\nabla_{\mu} \varepsilon = (\partial_{\mu} - \frac14 \omega_{\mu}^{ab} \gamma_{ab}) \varepsilon + \frac14 ( \mathcal{K}_i \partial_{\mu} z^i - \mathcal{K}_{\bar{\imath}}\partial_{\mu} \bar{z}^{\bar{\imath}} )\varepsilon + \ii g \xi_{\Lambda} A_{\mu}^{\Lambda} \varepsilon \, ,
\ee
and we have defined
\bea
 T^{-}_{\mu \nu} = 2 \ii \im \mathcal{N}_{\Lambda \Sigma} L^I F^{J-}_{\mu \nu} \, , \qquad
 G_{\mu \nu}^i = - g^{i \bar{j}} \bar{f}_{\bar{j}}^{\Lambda} \im \mathcal{N}_{\Lambda \Sigma} F_{\mu \nu}^{\Sigma -} \, ,
\eea
where $L^I$ are the upper part of the covariantly holomorphic section $\cV$ 
\be
\cV=  \left( \begin{array}{c} L^{\Lambda} \\ M_{\Lambda} \end{array} \right) \equiv e^{\mathcal{K}/2} \Omega = e^{\mathcal{K}/2}  \left( \begin{array}{c} X^{\Lambda} \\ F_{\Lambda} \end{array} \right) ,
\ee
$\cK$ is the K\"ahler potential, and $f^{\Lambda}_i$ are defined as
\be
 \left( \begin{array}{c} f^{\Lambda}_i \\ h_{\Sigma, i} \end{array} \right) \equiv \nabla_i \cV = \left( \partial_i + \frac12 \partial_i \mathcal{K} \right) \cV \, .
\ee
Further definitions can be found for instance in \cite{Andrianopoli:1996cm}. Finally  $[\gamma_a,\gamma_b]$ denotes the antisymmetrized product with unit weight, \ie\;$[\gamma_a,\gamma_b]  = \frac 1 2 ( \gamma_a \gamma_b - \gamma_b \gamma_a )$. We set the Fayet-Iliopoulos parameters $\xi_{\Lambda}=1$ for $\Lambda=0,...3$. For the four-dimensional solution described in section \ref{stu4D} (see \eqref{sol} and \eqref{gf}), we choose the following vierbeins
\be
 \begin{aligned}
  & e^0_t = \mathcal{H}(r)^{-1/4}\sqrt{f(r)} \, ,\qquad
  && e^1_r = \frac{ {\mathcal{H}(r)}^{1/4}}{\sqrt{f ( r)}} \, , \\
  & e^2_{\theta} = r  {\mathcal{H}(r)}^{1/4} \, ,
  && e^4_{\phi} = r {\mathcal{H}(r)}^{1/4} \sinh (\theta) \, ,
 \end{aligned}
\ee
and the non-vanishing components of the spin connection are then
\be
\omega_{t}^{01} = U'(r) \, , \qquad \omega_{\theta}^{12} = h'(r) \, , \qquad \omega_{\phi}^{13} =  h'(r) \sinh (\theta) \, , \qquad \omega_{\phi}^{23} = \cosh (\theta) \, .
\ee
Assuming a Killing spinor that fulfills the following relation 
\be
 \varepsilon = (a \ii \gamma_0 +  b \gamma_1) \varepsilon \, ,
\ee
with
\be
 a = \frac{\ii} {\sqrt{f ( r )}} \, , \qquad b = - \frac{2 g r}{ \sqrt{f ( r )}} \mathcal{H}^{1/2} \, ,
\ee
the supersymmetry equations \eqref{susyvar4D} simplify considerably, and one obtains the following explicit solution for the Killing spinor
\be
 \varepsilon = \frac{1}{2 \sqrt{g r}} e^{\frac{t}{2n}} \mathcal{H}^{-1/8}
 e^{-\frac12 \gamma_{012} \theta} e^{-\frac12 \gamma_{23} \phi}
 \left( \sqrt{f(r)-\ii} - i \gamma_1 \sqrt{f(r)+\ii} \right) (1-\gamma_0) \varepsilon_0 \, ,
\ee
where $\epsilon_0$ is an arbitrary spinor in four dimensions.
Notice that in the limit of constant scalars, namely $b_1=b_2=b_3=b_4=Q$ we recover the Killing spinor of \cite{Romans:1991nq}.

\subsection{Six dimensions}
\label{6D:BPS:proof}

In this section we explicitly construct the Killing spinor from the BPS equations of six-dimensional F(4) Romans gauged supergravity coupled to one vector multiplet.
We mostly follow the conventions of \cite{Hosseini:2018usu}, briefly recapping only the quantities relevant in our case, and referring to that paper for the details we omitted here for brevity. The supersymmetry variations of the fermions are
\bea
 \label{susyeqn6D}
 \delta \psi_{A \mu} & = \nabla_{\mu} \varepsilon_A -\frac{1}{2} g \sigma_{AC}^x A_{x \mu} \varepsilon^C  + \frac1{16} e^{-\sigma} [\widetilde{T}_{[AB] \nu \lambda} \gamma_7 - T_{(AB) \nu \lambda}]
 (\gamma_{\mu}^{\nu \lambda} - 6 \delta_{\mu}^{\nu} \gamma^{\lambda}) \varepsilon^B + S_{AC}\gamma^{\mu} \varepsilon^C \\
 & + \frac{\ii}{32}e^{2\sigma} H_{\nu \lambda \rho} \gamma_7 (\gamma_{\mu}^{\mu \lambda \rho} - 3 \delta_{\mu}^{\nu} \gamma^{\lambda \rho}) \varepsilon_A \, , \\ 
 \delta \lambda^I_A & = \ii P^{I}_{x \mu} \sigma^{x}_{AC} \gamma^{\mu} \varepsilon^C - \ii P^I_{0 \nu} \epsilon_{AC} \gamma^7 \gamma^{\nu} \varepsilon^C + \frac{\ii}{2} e^{-\sigma} T_{\nu \rho}^I \gamma^{\nu \rho} \varepsilon_A + M_{AB}^I \varepsilon^B \, , \\
\delta \chi_A & = \frac{\ii}{2} \gamma^{\nu} \partial_{\nu} \sigma \varepsilon_A + \frac{\ii}{16} e^{-\sigma} [\widetilde{T}_{[AB] \nu \lambda} \gamma_7 - T_{(AB) \nu \lambda}]  \gamma^{\nu \lambda} \varepsilon^B + \frac{1}{32} e^{2\sigma} H_{\nu \lambda \rho} \gamma_7 \gamma^{\nu \lambda \rho} \varepsilon_A + N_{AB} \varepsilon^B \, ,
\eea
where the capital Greek indices are raised and lowered with the $\SO(4,n_{\text{V}})$ invariant metric and the indices $A,B,\ldots$ with the antisymmetric tensor $\epsilon_{AB}$. The objects appearing in the susy equations are defined as
\be
 \widetilde{T}_{[AB] \nu \lambda}  = \epsilon_{AB} L_{0 \Sigma}^{-1} \hat{F}_{\nu \lambda }^{\Sigma} \, , \qquad
 T_{(AB) \nu \lambda} =  \sigma_{AB}^x L_{x \Sigma}^{-1} F_{\nu \lambda }^{\Sigma} \, , \qquad
 T_{I \nu \lambda} = L_{I \Sigma}^{-1} \hat{F}_{\nu \lambda }^{\Sigma} \, ,
\ee
and the matrices $N_{AB}$, $S_{AB}$, $M^I_{AB}$, along with a convenient parameterization of the scalar coset ${L_{\Lambda}}^{\Sigma}$  are defined in \cite{Hosseini:2018usu}, to which we refer for all missing definitions. In our case they boil down to
\bea
N_{AB} & = \frac14 (g \cosh(\phi_3) e^{\sigma} -3m e^{-3\sigma}) \epsilon_{AB} \, , \\
S_{AB} & = \frac{\ii}{4} (g \cosh(\phi_3) e^{\sigma} + m e^{-3 \sigma})  \epsilon_{AB} \, ,\\
M_{AB} & = - 2 g \sinh(\phi_3) e^{\sigma} \sigma^3_{AB} \, .
\eea

As in \cite{Alday:2014fsa}, also in our case the only component of the $\SU(2)$ gauge field is the one in the $i=3$ direction.
With reference to \eqref{6D_warp} we choose the following vielbeins
\be
 \begin{aligned}
  & e_r^0 = \frac{\mathcal{H}^{1/8} }{f(r)^{1/2}} \, , \qquad  e_t^1 = \frac{3}{\sqrt{2}} \frac{f(r)^{1/2}}{\mathcal{H}^{3/8}} \, , \qquad \quad e_{\chi}^2 = r \mathcal{H}^{1/8} \, , \qquad  e_{\theta}^3  = r \mathcal{H}^{1/8}  \sinh(\chi) \, , \\
  & e_{\psi}^4 =  r \mathcal{H}^{1/8} \sinh(\chi) \sin ( \theta) \, , \qquad e_{\phi}^5 =  r \mathcal{H}^{1/8}\sinh(\chi) \sin(\theta) \sin(\chi) \, ,
 \end{aligned}
\ee
The non-vanishing components of the spin connection read
\be
 \begin{aligned}
 & w^{01}_t  =  \frac{U'(r) \sqrt{V(r)}}{2 \sqrt{U(r)}} \, , \qquad
 w_{\chi}^{02} = - \frac{h'(r) \sqrt{V(r)}}{2  \sqrt{h(r)}} \, , \qquad
 w^{03}_{\theta} = - \frac{\sinh(\chi) h'(r) \sqrt{V(r)}}{2 \sqrt{h(r)}} \, , \\
 & w^{23}_{\theta}  = - \cosh(\chi) \, , \qquad \qquad \,
 w^{04}_{\psi} = - \frac{\sinh(\chi) h'(r) \sqrt{V(r)} \sin \theta}{2 \sqrt{h(r)}} \, , \qquad
 w^{24}_{\psi} = - \cosh(\chi) \sin( \theta) \, , \\
 & w^{34}_{\psi}  =- \cos (\theta) \, , \qquad
 w^{05}_{\phi} = - \frac{\sinh(\chi) h'(r) \sqrt{V(r)} \sin(\theta) \sin(\psi)}{2 \sqrt{h(r)}} \, , \\
 & w_{\phi}^{25} = - \cosh(\chi) \sin(\theta)\, \sin(\psi) , \qquad w^{35}_{\phi} =  - \cos (\theta) \, \sin(\psi) \, , \qquad  w^{45}_{\phi} =  - \cos (\psi) \, .
 \end{aligned}
\ee

We are going to consider first the variation of the dilatino $\chi_{A}$. Given our truncation, we can see that imposing the relation
\be \label{proj2}
 \left( {\delta_A}^B +\ii x(r) \gamma^{0}  \sigma_{AC}^x\epsilon^{CB} +y(r) \gamma^1 {\delta_A}^B  \right) \varepsilon_B =0 \, ,
\ee
with
\bea
& x(r)  = - \frac{3 \ii r^2}{\sqrt{2 b_1 \left(b_2+r^3\right)+2 b_2 r^3 +2 r^6-9r^4}} \, , \\
& y(r) = - \frac{\sqrt2 \sqrt{b_1+r^3} \sqrt{b_2+r^3}}{\sqrt{2 b_1 \left(b_2+r^3\right)+r^3 \left(2 b_2+2 r^3-9 r\right)}} \, ,
\eea
where $x(r)^2+y(r)^2 =1$, the gravitino equation reduces to
\be
 \begin{aligned}
  \delta \psi_{A,t} &= \left( \partial_t -\, \frac{1}{2}  \left(  1-\ii (\phi_1 +\phi_2) \right) \, {m_{A}}^B  \, \right) \varepsilon_B = \left( \partial_t -\, \frac{1}{2n} \, {m_{A}}^B  \, \right) \varepsilon_B \, , \\ \label{radial}
  \delta \psi_{A,r} &= \left( \partial_r + f_1(r) {m_{A}}^B \gamma^0 + f_{2}(r) {\delta_{A}}^B \right) \epsilon_B \, , \\
  \delta \psi_{A,\chi} &= \left( \partial_{\chi}  -\frac12 \gamma_{012} {m_{A}}^B  \right) \varepsilon_B \, , \\
  \delta \psi_{A,\theta} &= \left(  \partial_{\theta}  - \frac12 \sinh(\chi)  \gamma_{013}{m_{A}}^B   -\frac12 \cosh(\chi) \gamma_{23} \right) \varepsilon_B \, , \\
  \delta \psi_{A,\psi} &= \left( \partial_{\psi} -\frac12 \sinh(\chi) \sin(\theta) \gamma_{014}{m_{A}}^B -\frac12 \cosh(\chi) \sin(\theta) \gamma_{24} -\frac12 \cos{\theta} \gamma_{34} \right) \varepsilon_B \, , \\
  \delta \psi_{A,\phi} &= \bigg( \partial_{\phi}-\frac{1}{2} \sin (\theta ) \sinh (\chi) \sin (\psi ) \gamma_{015} {m_{A}}^B +\frac{1}{2} \sin (\theta ) \cosh (\chi) \sin (\psi ) \gamma_{25} \\
  &- \frac{1}{2} \cos (\theta ) \sin (\psi ) \gamma_{35} -\frac{1}{2} \cos (\psi)  \gamma_{45} \bigg) \varepsilon_B \, .
 \end{aligned}
\ee
Here, we defined the functions $f_1(r)$ and $f_2(r)$ as
\bea
f_1(r) & = \frac{9 \left( b_1 \left(2 b_2+r^3\right)+ b_2 r^3\right)}{16 r \left( b_1+r^3\right) \left(b_2+r^3\right)} \, , \\
f_2(r) & = \frac{ \left(r^3 (b_1+b_2)+4 b_1 b_2-2 r^6\right)}{2 \sqrt{2} r \sqrt{b_1+r^3} \sqrt{b_2+r^3} \sqrt{2 r^3 (b_1+b_2)+2 b_1 b_2+2 r^6-9 r^4}} \, .
\eea
and we defined ${m_{A}}^B = \sigma_{AC}^x\epsilon^{CB}$. The $t,\chi,\theta, \psi,\phi$ equation can be solved immediately by the following \cite{Lu:1998nu}:
\be \label{KStotal}
\varepsilon_A (r,t,\chi,\theta, \psi, \phi) = e^{\frac{t}{2n} {m_{A}}^B } e^{ \frac{ \chi}{2} \gamma_{012} {m_{A}}^B} e^{\frac{\theta}{2} \gamma_{23}} e^{\frac{\psi}{2} \gamma_{34}} e^{\frac{\phi}{2} \gamma_{45}} \varepsilon_{A}(r) \, ,
\ee
and the radial component of \eqref{radial}, together with the relation \eqref{proj2} can be solved by standard methods of \cite{Romans:1991nq}, resulting in 
\be \label{rad}
\varepsilon_{A}(r) = (u(r) + v(r) \gamma^1) (\delta_{AB}- \ii \bar{\Gamma}_{AB}) \varepsilon_{0,B} \, .
\ee
Here $\varepsilon_{0,B}$ is a doublet of constant spinors, $\bar{\Gamma}_{AB} = \gamma_0 \sigma_{AC}^3 \epsilon^{CB}$ and
\bea
u(r) & = \sqrt{\frac{1+x(r)}{y(r)}} e^{w(r)} = \sqrt{ \frac{\sqrt{2 b_1 \left(b_2+r^3\right)+r^3 \left(2 b_2+2 r^3-9 r\right)}+3 \ii r^2}{\sqrt2 \sqrt{b_1+r^3} \sqrt{b_2+ r^3}}} e^{w(r)} \, , \\
v(r) & = - \sqrt{\frac{1-x(r)}{y(r)}} e^{w(r)} =  - \sqrt{\frac{\sqrt{2 b_1 \left( b_2+r^3\right)+r^3 \left(2 b_2+2 r^3-9 r\right)}-3 \ii r^2} { \sqrt2  \sqrt{b_1+r^3} \sqrt{ b_2+ r^3} } }e^{w(r)} \, ,
\eea
with
\be
w(r) = \int^r f_1(r') dr' = \frac{9}{16} \left(-\frac{1}{3} \log \left(b_1+r^3\right)-\frac{1}{3} \log \left(b_2+r^3\right)+2 \log (r)\right) ,
\ee
hence
\be
e^{w(r)} = \frac{r^{9/8}}{\left(b_1+r^3\right)^{3/16} \left(b_2+r^3\right)^{3/16}} \, .
\ee
The total Killing spinor is then given by combining \eqref{KStotal} with the radial dependent part in \eqref{rad}.
Notice that the second bracket of \eqref{rad} projects out half of the supersymmetries, which signals the fact that the solution indeed is 1/2 BPS. It is easy to check that this expression also solves the gaugino equation $ \delta \lambda^I_A =0$ in \eqref{susyeqn6D}.

\section{On-shell action via holographic renormalization \label{AppB}}

In this section we compute the renormalized on-shell action in the grand-canonical ensemble for the solutions we described in the main sections.
In \cite{Batrachenko:2004fd} the on-shell action for the corresponding $\mathcal{N} =2$ four-dimensional gauged supergravity and Romans $F(4)$ spherical solutions was computed. Generalizing the computation to hyperbolic horizons of different topology requires a minimal modification of their procedure, which we explain in this appendix, following their notation closely. Other relevant references for holographic renormalization in this context are for instance \cite{Papadimitriou:2011qb,An:2017ihs,Cabo-Bizet:2017xdr} and we will make use of them when deriving the counterterms. Notice that here $d$ denotes the dimension of the boundary.

For both setups the action can be cast in the following form (see \cite[(5.1)]{Chow:2011fh}):
\begin{eqnarray} \label{action1}
S = &  -& \frac{1}{16 \pi G_{d+1}} \int_M \rd^{d+1} x \sqrt{-g} \left( R-\frac12 G_{ij} \partial_{\mu} z^i \partial^{\mu} z^j -\frac14 M_{IJ} F_{\mu \nu}^I F^{\mu \nu, J} - V(z^i) \right) \nonumber \\
& + & \frac{1}{8 \pi G_{d+1}} \int_{\partial M} \rd^{d}x \sqrt{-h} \Theta \, ,
\end{eqnarray}
where $M$ is a $(d+1)$-dimensional spacetime with metric $g_{\mu \nu}$, boundary $\partial M$ with induced metric $h_{\mu \nu}$. In this case $\Theta$ is the trace of the extrinsic curvature $\Theta_{\mu \nu}$ of the boundary $\Theta_{\mu \nu} = -\frac12 (\nabla_{\mu} \xi_{\nu} + \nabla_{\nu} \xi_{\mu})$, where $\xi^{\mu}$ is the outward-pointing normal to $\partial M$. 

We can massage the bulk term of the action \eqref{action1} by making use of the trace of the Einstein's equation, to obtain 
\be \label{I1}
I_{\text{bulk}} = -\frac{1}{16 \pi G_{d+1}} \int_{M} \rd^{d+1} x \sqrt{-g} \left[ - \frac{1}{2(d-1)} M_{IJ} F_{\mu \nu}^I F^{\mu \nu, J} + \frac{2}{d-1} V(z^i) \right] .
\ee
This latter expression can be rewritten as \cite{Batrachenko:2004fd} 
\be \label{Rphiphi}
 I_{\text{bulk}} = -\frac{1}{8 \pi G_{d+1}} \int_{M} \rd^{d+1}x \sqrt{-g} {R_{\phi}}^{\phi} \, .
\ee
We use an ansatz for the metric of the form 
\be \label{ansatz_metricd}
\rd s^2 = - \frac{\mathcal{H}^{-(d-2)/(d-1)}f(r)}{ \textbf{c}^2} \rd t^2 + \frac{\mathcal{H}^{1/(d-1)}}{f(r)} \rd r^2+ \mathcal{H}^{1/(d-1)} r^2 \rd s_{\mathbb{H}^{{d-1}}}^2 \, ,
\ee
with
\be
f(r) = -1 -\frac{\mu}{r^{d-2}} + \tilde{g} r^2 \mathcal{H}(r)^2 \, , \qquad \mathcal{H} = \prod_I H_I \, ,
\ee
which encompasses both the four-dimensional configurations of section \ref{stu4D} and those of section \ref{6dF4} for a suitable choice of $\textbf{c}$ and $\tilde{g}$.
A direct computation of the term in \eqref{Rphiphi}, once we define $B(r) = \frac{1}{2(d-1)} \log \mathcal{H}(r)$, gives
\be
 \sqrt{- g} {R_{\phi}}^{\phi} = -\frac{1}{\textbf{c}} \frac{\rd}{\rd r} \left( B'(r) r^{d-1} f(r)+ r^{d-2} (f(r)+1) \right) .
\ee
This term differs from the spherical case treated in \cite{Batrachenko:2004fd} by the sign of the last addendum.
The bulk term therefore yields
\be
 I_{\text{bulk}}= \frac{\Vol(\mathbb{H}^{d-1}) \beta}{16 \pi \textbf{c} G_{d+1}} \left(B'(r_{\text{inf}})  r_{\text{inf}}^{d-1} f(r_{\text{inf}}) + r_{\text{inf}}^{d-2} (f(r_{\text{inf}}) +1) - r_+^{d-2} \right) ,
\ee
where we used that $f(r_+) =0$.
As for the Gibbons-Hawking term, the normal outward pointing is given by $n^r = \sqrt{f(r)} \mathcal{H}^{-1/(2(d-1))} = \sqrt{f(r)} e^{-B(r)}$.
The extrinsic curvature reads
\be
 \Theta = -\frac{e^{-B(r)} \left(2 f(r) \left(r B'(r)+d-1\right)+r f'(r)\right)}{2 r \sqrt{f(r)}} \, ,
\ee
which yields a Gibbons-Hawking term of the form
 \be
 I_{\text{GH}} = - \frac{\Vol(\mathbb{H}^{d-1}) \beta}{8 \pi \textbf{c} G_{d+1}}  r_{\text{inf}}^{d-2} \left[ f(r) \left(r B'(r)+d-1\right)+ \frac12 r f'(r)\right]_{r=r_{\text{inf}}} \, .
 \ee
Here we used
\be
 \sqrt{-h} = \frac{e^{B(r)}}{\textbf{c}}r^{d-1} \sqrt{f(r)}  v_{d-1}  \, .
\ee
where $v_2 =\sinh (\theta) $ and $v_4= \sinh^3 (\chi) \sin^2(\theta)\sin(\psi)$. This leads to
\bea
 \label{Ibulk4}
 I_{\text{bulk}}+I_{\text{GH}} & = \frac{ \Vol(\mathbb{H}^{d-1}) \beta}{8 \pi \textbf{c} G_{d+1}} \left(-(d-2) r_{\text{inf}}^{d-2} f(r_{\text{inf}}) -\frac12 r_{\text{inf}}^{d-1} f'(r_{\text{inf}}) +r_{\text{inf}}^{d-2}-r_+^{d-2} \right) .
\eea
We will now spell out the relevant counterterms for the different cases, specializing to the different $d+1=D=4$ and $D=6$ cases, dealing first with the former. 

\subsection*{Four dimensions}

The holographic renormalization procedure in $D=4$ follows from \cite{Cabo-Bizet:2017xdr}, where the counterterms for $\mathcal{N}=2$ $U(1)$-gauged supergravity coupled to three vector multiplets were derived. In particular our solution is purely electric hence the counterterms boil down to
\be
 I_{\text{ct}}= \frac{1}{8 \pi G_4} \int \rd^3x \sqrt{-h} \left( \mathcal{W}(z^i) +\frac1{2 \tilde{g}} \mathcal{R}_3+\ldots \right) ,
\ee
where the ellipsis denotes the terms which are subleading once the cutoff is send to infinity. $\cal{R}$ is the Ricci scalar of the boundary, and $\mathcal{W}$ is a function of the scalar fields called superpotential. We have the following expression for the Ricci scalar of the boundary $\mathcal{R}_3$:
\be
 \mathcal{R}_3 = - \frac{2}{r^2} e^{-2 B(r)} \, .
 \ee
and the superpotential that drives the flow is given by \cite{Gnecchi:2014cqa}
 \be
 \mathcal{W} = \frac{\tilde{g}}{2} \sum_{I = 0}^3 X^I \, ,
\ee
which coincides with that used in \cite{Batrachenko:2004fd,Cabo-Bizet:2017xdr,Cassani:2019mms} and amounts to imposing Neumann boundary conditions on the scalar fields, a procedure that is compatible with supersymmetry \cite{Freedman:2013ryh}. Adding this to the action \eqref{Ibulk4} we finally find
\be
 I_{\text{tot}} = I_{\text{bulk}} +I_{\text{GH}}+ I_{\text{ct}} = - \frac{\beta \? \Vol(\mathbb{H}^{2}) }{8 \pi \textbf{c} G_4} \left(r_+ + \frac{\mu}2 \right) ,
\ee
which indeed reduces to $I_{\text{tot}} =  - \frac{\beta \? \Vol(\mathbb{H}^{2}) }{8 \pi  \textbf{c} G_4 }r_+$ in the BPS limit.

\subsection*{Six dimensions}
In $D=6$ we have the following counterterms  \cite{Batrachenko:2004fd}:
\be \label{ct6}
 I_{\text{ct}}= \frac{1}{8 \pi G_d} \int \rd^3x \sqrt{-h} \left( \mathcal{W}(z^i) +\frac1{6g} \mathcal{R}_5+ \frac{1}{18g^3} \left( \mathcal{R}_{5,ab} \mathcal{R}_5^{ab} - \frac{5}{16} \mathcal{R}_5^2\right) \right) ,
\ee
where $\mathcal{R}_5$ and $\mathcal{R}_{5,ab}$ are respectively the Ricci scalar and the Ricci tensor of the five-dimensional boundary metric\footnote{A full treatment of the supersymmetric boundary counterterms for matter coupled $D=6$ supergravity to our knowledge is still unknown (see \cite{An:2017ihs} for a treatment for $D=5$). Nevertheless the scalar and vector falloff is very rapid at infinity, so that there is no contribution from the matter fields for our configurations. See also the discussion later in the text.}. Notice that the terms of higher power in the curvature this time contribute to the free energy one the cutoff is removed. We give here the form for the Ricci scalar:
\be 
 \mathcal{R}_5 = - \frac{12}{r^2} e^{-2 B(r)} \, .
 \ee
 In our case one could for instance choose as counterterm the superpotential $\mathcal{W}$ appearing in the susy variations, $S_{AB} = \mathcal{W} \epsilon_{AB}$, which shows the same falloff behaviour and reduces to that of \cite{Batrachenko:2004fd} for $X^1 = X^2$. For the six-dimensional configurations taken into consideration, however, the asymptotic falloff of the scalars is very rapid. The expansion of the superpotential $\mathcal{W}$ contains terms which are at least quadratic in the fields (see for instance \cite{Papadimitriou:2006dr}) therefore it turns out that the scalars do not contribute to the boundary counterterm,%
\footnote{The scalar behaviour at infinity is $z^i \sim \text{const} + \cO(r^{-3})$, while $\sqrt{-h} \sim r^5$. Indeed, also for the known cases \cite{Batrachenko:2004fd,Alday:2014fsa} the counterterm contribution is just a constant independent of the scalars falloff, $\mathcal{W} = 4\tilde{g}$.}
and indeed a term $\mathcal{W} = 4\tilde{g}$ suffices to renormalize the on-shell action. Putting together expressions \eqref{Ibulk4} and \eqref{ct6}, we get
\be
 I_{\text{tot}} = I_{\text{bulk}} + I_{\text{GH}}+ I_{\text{ct}} = - \frac{\beta \? \Vol(\mathbb{H}^{4}) }{8\pi  \textbf{c} G_6} \left( r_+^3 +\frac{\mu}2 \right) ,
\ee
which indeed reduces to $I_{\text{tot}} = - \frac{\beta \? \Vol(\mathbb{H}^{4}) }{8\pi  \textbf{c} G_6}  r_+^3 $ in the BPS limit.

\subsection*{Thermodynamics relation and conserved charges in $d$ dimensions}

In this section we prove the formula
\be \label{W}
W = \frac{I}{\beta} = E- T S_{\text{BH}} - \phi^I Q_I \, ,
\ee
again generalizing the computation of \cite{Batrachenko:2004fd} to the hyperbolic case, following closely their notation. We start from \eqref{I1} and we rewrite it with the help of the $R_{tt}$ component of the Einstein's equations, assuming that all matter fields are independent of time, as it is the case for the solutions considered in this paper. We obtain
\be
R_t^t = \frac12 M_{IJ} F_{tr}^I F^{tr,J} -\frac1{4(d-1)} M_{IJ}F_{\mu \nu}^I F^{\mu \nu,J} +\frac{1}{d-1}V\,,
\ee
hence \eqref{I1} becomes
\be \label{startI}
I_{\text{bulk}} =-\frac{1}{8\pi G_{d+1}} \int \rd^{d+1} x \sqrt{-g} \left( R_t^t -\frac{1}{2}M_{IJ} F_{tr}^IF^{J,tr} \right) \,.
\ee
We have verified that for the metric of the form \eqref{ansatz_metricd} the following holds
\be
R_t^t = \frac{1}{\textbf{c} \sqrt{-g}} \frac{\rd}{\rd r}\left( \sqrt{-h} \Theta_t^t \right) \,.
\ee
Moreover, we have the Maxwell's equation $\partial_r (\sqrt{-g} M_{IJ} F^{J, rt})=0$. We define the following conserved charges $\mathbf{q}_I$ as
\be
\mathbf{q}_I = \frac{\sqrt{-g}}{v_{d-1}} M_{IJ} F^{J,rt}\,.
\ee
Plugging these expressions into \eqref{startI} we arrive at the following formula for $I_{\text{bulk}}$:
\bea
I_{\text{bulk}} & = -\frac{1}{8 \pi G_{d+1}} \int \rd^{d}x \int_{r_+}^{r_{\text{inf}}} \rd r \frac{\rd}{\rd r} \left( \frac12 A_t^I \mathbf{q}_I  + \frac{\sqrt{-h}}{\textbf{c}} \Theta_t^t \right) \\
& = -\frac{\beta\Vol(\mathbb{H}^{d-1}) }{8 \pi G_{d+1}} \left(\frac12 A_t^I \mathbf{q}_I  + \frac{\sqrt{-h}}{v_{d-1} \textbf{c}} \Theta_t^t \right) \bigg|_{r_+}^{r_{\text{inf}}} \,.
\eea
To regularize the action we need to add the Gibbons-Hawking term, therefore, the full regularized action reads
\be \label{iregthermo}
I_{\text{reg}} = \beta W = \frac{\beta\Vol(\mathbb{H}^{d-1})}{8 \pi G_{d+1}} \left(- \frac12 \phi^I \mathbf{q}_I  + \frac{\sqrt{-h}}{v_{d-1} \textbf{c}}(\Theta -\Theta_t^t) \Big|_{r_{\text{inf}}} + \frac{\sqrt{-h}}{v_{d-1} \textbf{c}} \Theta_t^t \Big|_{r_+} \right) \,.
\ee
The first term gives directly the product of the chemical potentials, defined as
\be
\phi^I= A^t(r_{\text{inf}})- A^t(r_+)\,,
\ee
 and the electric charges, once we define the charges as $Q_I = \frac{\Vol(\mathbb{H}^{d-1})}{16 \pi G_{d+1}} \mathbf{q}_I$.
We see that the second term is related to the ADM mass of the system, while the third one is related to the product of the temperature $T$ and the Bekenstein-Hawking entropy $S_{\text{BH}}$.
Let us focus on the latter. Given the definitions
\be
 T = \frac{1}{4\pi \mathbf{c} \sqrt{ \mathcal{H}(r_+)}}  \frac{\rd f}{\rd r}\bigg|_{r_+} \, , \qquad S_{\text{BH}} = \frac{A}{4G_{d+1}} = \frac{\Vol(\mathbb{H}^{d-1})}{8 \pi G_{d+1}} \big( 2\pi \sqrt{\mathcal{H}(r_+)} r_+^{d-1} \big) \, ,
\ee
we obtain
\be
TS_{\text{BH}} = - \frac{\Vol(\mathbb{H}^{d-1})}{8 \mathbf{c} \pi G_{d+1}} \frac{\sqrt{-h}}{v_{d-1}} \Theta_t^t \bigg|_{r_+} \,,
\ee
which holds for a metric of the form \eqref{ansatz_metricd}. 

The energy is extracted from the renormalized boundary stress energy tensor $T^{ab} = \frac{2}{\sqrt{-h}} \frac{\delta I}{\delta h_{ab}}$ in this way:
\be
E = \frac{1}{8\pi G_{d+1}} \int_{\Sigma} \sqrt{\sigma} u_a T^{ab} K_b \,,
\ee
where $K_a$ is the Killing vector field associated with an isometry of the boundary induced metric (in this case, time translations). $\Sigma$ is the spacelike section of the boundary, $\sigma_{ab}$ is the induced metric on $\Sigma$, and $u^a = \sqrt{-h^{tt}}(1,0,0)$ is the unit normal vector to $\Sigma$.
We will first compute the regulated energy $E_{\text{reg}}$, discussing the counterterms later. The regularized energy reads
\be
E_{\text{reg}}= \frac{\Vol(\mathbb{H}^{d-1})}{8 \pi G_{d+1}} \frac{\sqrt{-h}}{v_{d-1}\textbf{c}} (- \Theta_t^t +\Theta) \,.
\ee 
Plugging all these relations into \eqref{iregthermo} we get
\be
W_{\text{reg}} = E_{\text{reg}} - T S_{\text{BH}} - \phi^I Q_I \,.
\ee
This is the relation valid for the regularized quantities. The renormalized ones are obtained by adding the counterterms spelled out in the previous subsections. The counterterms contribute only to the renormalization of the mass, giving $E_{\text{ren}}$.
Hence, the thermodynamics relation \eqref{W} holds, as expected. 

For the records, we report here the explicit values of the energy $E_{\text{ren}} =E$ (black hole mass), entropy $S_{\text{BH}}$ and charges $Q^I$ for the solutions considered in the main text. For the four-dimensional solutions in section \ref{stu4D} we have
\be \label{charges_sugra}
 \begin{aligned}
  E & = \frac{\Vol(\mathbb{H}^{2})}{16 \pi G_{4}} \left( \mu - \frac12(b_1+b_2+b_3+b_4) \right) , \\
  S_{\text{BH}} & = \frac14 \sqrt{H_1(r_+) H_2(r_+) H_3(r_+) H_4(r_+)} \?r_+^2 \, ,\\
  Q_I & =  \ii b_I \frac{\Vol(\mathbb{H}^{2})}{16 \pi G_{4}} \, , \qquad I = 1, \ldots, 4 \, .
 \end{aligned}
\ee
For the six-dimensional solutions of section \ref{6dF4} we find
\be
\begin{aligned}
 E & =  \frac{3\Vol(\mathbb{H}^{4})}{8 \pi \sqrt2 G_{6}} \left(\mu-\frac34 (b_1+b_2) \right) , \\
 S_{\text{BH}} & = \frac14 \sqrt{H_1(r_+) H_2(r_+)} \?r_+^4\,, \\
 Q_I & =  \ii \frac{3}{\sqrt2}b_I \frac{\Vol(\mathbb{H}^{4})}{16 \pi G_{4}} \, , \qquad I = 1, 2 \, .
\end{aligned}
\ee

\section{Computation of the charges}
\label{AppC}

We now demonstrate that the black hole charges computed from supergravity match those computed in the SCFT. We do so only for the ABJM model. 

The trace representation in \eqref{trace_representation} contains three independent flavor charges which correspond to some choice of basis for chemical potentials, represented by flavor gauge fields, satisfying the constraint \eqref{chemical_potential_constraint}. In order to compare with the bulk charges, it is useful to implement the constraint using a Lagrange multiplier charge $\Lambda$ 
\be
 \label{Zn:Tr:Lambda}
Z^{\text{susy}}_{n}=\text{Tr}_{\mathbb{H}^{d-1}}e^{-2\pi n\left(H- \ii \sum_{I}\alpha^{I}Q_{I} + \ii \frac{n-1}{n}Q_{R}- \ii \Lambda\sum_{I}\alpha^{I}\right)} \, .
\ee
From this expression, we can calculate the following
\be
\partial_{n}\left(-\log Z^{\text{susy}}_{n}\right)=2\pi H-\ii\sum_{I}\alpha^{I}Q_{I}+\ii Q_{R} - \ii \Lambda\sum_{I}\alpha^{I} \, ,
\ee
and
\be
\frac{1}{2\pi n}\partial_{\alpha^{I}}\left(-\log Z^{\text{susy}}_{n} \right)=-\ii ( Q_{I} +  \Lambda ) \, .
\ee
In order to compare with the bulk, we first recast \eqref{Zn:Tr:Lambda} in terms of the bulk quantities $I,\beta$, and $\phi^I$:
\be
I=-\log\left[\text{Tr}_{\mathbb{H}^{d-1}}e^{-\beta\left(E - \sum_{I}\phi^{I}Q_{I}+ \ii \frac{2\pi-\beta}{4\beta}\sum_{I}Q_{I}+\ii\frac{\beta-2\pi}{\beta}Q_{R} - \Lambda\sum_{I}\phi^{I}+\ii\frac{2\pi-\beta}{\beta}\Lambda\right)}\right].
\ee
As a check on this expression, the constraint charge indeed now imposes
\be
\sum_{I = 1}^{4} \phi^{I} = \ii \frac{2\pi-\beta}{\beta}= \ii \frac{1-n}{n} \, ,
\ee
that is \eqref{constr_chempot}. 

In terms of bulk variables, we can now calculate
\be
-\frac{1}{\beta}\partial_{\phi^{I}}I \Big|_{\beta} = Q_{I}+\Lambda ,
\ee
which, from the definition \eqref{rel1}, implies,
\be
\mathcal{Q}_{I} = Q_{I}+\Lambda.
\ee
We also find that 
\be
\partial_{\beta}I\Big|_{\phi^{I}}  =E+\sum_{I}\phi^{I}\mathcal{Q}_{I}-\ii\Lambda-\frac{\ii}{4}\sum_{I}Q_{I}+\ii Q_{R} \, ,
\ee
yielding
\be
E=\partial_{\beta}I \Big|_{\phi^{I}}-\frac{1}{\beta}\sum_{I}\phi^{I}\partial_{\phi^{I}}I\Big|_{\beta}+\ii \Big(-Q_{R}+\frac{1}{4}\sum_{I}Q_{I} \Big)+\ii\Lambda \, ,
\ee
which is compatible with \eqref{rel1} only if we set 
\be
Q_{R}=\Lambda+\frac{1}{4}\sum_{I}Q_{I}=\frac{1}{4}\sum_{I}\mathcal{Q}_{I} \, .
\ee
 
We expect the subleading terms of the bulk gauge fields to capture the vacuum expectation value of the charges, \ie\,
\be
Q_{I}=\mathcal{N}b_{I},
\ee
where $\mathcal{N}$ is a normalization constant. We cannot extract all the $\mathcal{Q}_{I}$, because we do not know
$Q_{R}$ from the field theory. However, we do expect the following
equations for one less variable
\be
\mathcal{Q}_I-\frac{1}{4}\sum_{J}\mathcal{Q}_J = \ii \mathcal{N} \Big(b_{I}-\frac{1}{4}\sum_{J}b_{J} \Big).
\ee
Recall that $\cQ_I = -\frac{1}{\beta}\partial_{\phi^{I}}I \Big|_{\beta}$, see \eqref{rel1}.
One may now check, using the expressions \eqref{potential} and \eqref{Omega} for $b_I$ and $I$ as a function of the $\phi^I$, that this indeed holds, with 
\be
\mathcal{N} = \ii \frac{\Vol(\mathbb{H}^{2})}{8\pi \textbf{c} G_{4}},
\ee
after imposing \eqref{constr_chempot}. Therefore setting $\textbf{c}=2$ the value of the charges coincide with the supergravity ones \eqref{charges_sugra}.

\section{Rotating charged hyperbolic solutions \label{AppD}}

In this last section we take into consideration supersymmetric rotating black holes with hyperbolic event horizon that generalize the solutions of section \ref{warmup}\footnote{Hyperbolic rotating black holes with nontrivial scalar fields exist as well \cite{Hristov:2019mqp}, along with analogous magnetic configurations realizing the topological twist \cite{Klemm:2011xw,Hristov:2018spe} but we do not consider them here and we focus instead on the simple minimal gauged supergravity (``universal" truncation) solution.}. We compute their on-shell action and we show that it assumes a simple form, once the BPS constraints are enforced. We will make contact with the limiting procedure of \cite{Cabo-Bizet:2018ehj}, which allows to approach an extremal BPS limit in the complexified solution. 

The Kerr-Newman hyperbolic solution with purely electric charge reads:
\be
\rd s^2= - \frac{\Delta_r}{\Xi^2 \rho^2} (\rd t+a \sinh^2 (\theta) \rd \phi)^2 +\frac{\rho^2}{\Delta_r} \rd r^2 + \frac{\rho^2}{\Delta_{\theta}} \rd \theta^2 + \frac{\Delta_{\theta} \sinh^2 (\theta)}{\Xi^2 \rho^2} (a \rd t - (r^2+a^2)  \rd \phi)^2 \, ,
\ee
with 
\bea
 \label{warps}
 & \Delta_r = (r^2 +a^2) \bigg( \frac{r^2}{l^2} - 1 \bigg) - 2 m r +Q^2 \, , \\
 & \rho^2 = r^2 + a^2 \cosh^2 (\theta) \, , \qquad \Delta_{\theta} = 1+\frac{a^2}{l^2} \cosh^2 (\theta) \, , \qquad \Xi = 1+ \frac{a^2}{l^2} \, ,
\eea
and the gauge field
\be
A = - \frac{Q r}{ \Xi \rho^2} (\rd t + a \sinh^2 (\theta) \rd \phi) \, .
\ee
The on-shell action satisfies the thermodynamics relation (we set $G_4=1$)
\be \label{thermo_eq}
I = \beta(M- T S_{\text{BH}} - \phi_e Q_e - \Omega J) \, ,
\ee
where 
\bea
 \label{eq:thermo:rot:hyper:BH}
 & M = \frac{m}{(1 + \frac{a^2}{l^2}) }  \, , \qquad S_{\text{BH}} = 4 \pi \frac{r_+ +a^2}{\left( 1+\frac{a^2}{l^2} \right)} \, , \qquad J = \frac{a m}{(1 + \frac{a^2}{l^2}) } \, , \\
 & \phi_e= \frac{Q r_+}{r_+^2 +a^2} \, , \qquad Q_e = \frac{Q}{\left(1+ \frac{a^2}{l^2} \right)} \, , \qquad \Omega = \frac{a(l^2-r_+^2)}{l^2(a^2 +r_+^2)} \, , 
\eea
and
\be
\beta = \frac{4 \pi (r_+^2 +a^2)}{r_+ \left(-1 + \frac{a^2}{l^2} + \frac{3r_+^2}{l^2}- \frac{(Q^2-a^2)}{r_+^2} \right)} \, .
\ee
One can see that the boundary of spacetime takes the form
\be
\rd s^2 = -\frac{\rd t^2}{\Xi^2} + \frac{l^2 \rd \theta^2}{\Delta_{\theta}} + \frac{l^2}{\Xi} \sinh^2 (\theta) \rd \phi^2 \, ,
\ee
that can be cast in
\be \label{RH2}
\rd s^2 = \frac{\Delta_{\theta}}{\Xi^2} \left(- \rd \tau^2 +l^2(\rd \Theta^2 +\sinh^2 (\Theta) \rd \Phi^2) \right) ,
\ee
via the change of coordinates \cite{Klemm:2014nka}
\be
\tau = \frac{t}{\Xi} \, , \qquad \cosh (\Theta) = \cosh (\theta) \sqrt{\frac{\Xi}{\Delta_{\theta}}} \, , \qquad \Phi = \phi - \frac{a t}{l^2 \Xi} \, .
\ee
The metric \eqref{RH2} describes a space which is conformal to (part of) $\mathbb{R} \times \mathbb{H}^2$. 

The BPS condition, which can be read off from \cite{Caldarelli:1998hg} is given by
\be \label{BPScc}
m^4+2\left(1-\frac{a^2}{l^2} \right) m^2 Q^2 +\left( 1+ \frac{a^2}{l^2}\right)^2 Q^4 =0 \, ,
\ee
which has no solution for real $Q$, as expected from the previous sections. However, it has solutions for \emph{imaginary} $Q$ and $a$, which, as stated before, makes sense if we have in mind to work with a Euclidean solution (obtained by Wick rotating $t \rightarrow -\ii \tau$), for which the gauge field and metric will then be real.

We define $a \rightarrow \ii j$ and $Q_e \rightarrow \ii q_e$ with $j$ and $q_e$ real, and we set $l=1$ for simplicity. We use the BPS condition \eqref{BPScc} written in function of the latter parameters, to read off the value of the mass
\be
m= (j+1)q_e \, ,
\ee
where we chose the positive branch for regularity. We plug this relation into $\Delta_r$ in \eqref{warps} to express the charge $q_e$ as a function of the outer radius $r_+$, using the fact that $\Delta_r (r_+) =0$: 
\be
 \label{eq:Q:rp}
q_e = -(j \pm r_+)(r_+ \pm1) \, .
\ee
Given these relations, the on-shell action \eqref{thermo_eq} assumes the simple form
\be \label{onsh-simpl}
I = \frac{\pi (r_+ \pm j)^2}{(j-1)(j+1\pm 2r_+)}\,.
\ee
In terms of the chemical potentials and $\beta$, the on-shell action reads 
\be I = \pm \ii \frac{\beta(\phi_e+\ii)^2}{2(\Omega+\ii)}.\ee 
Notice that the chemical potentials $\Omega$ and $\phi_e$ satisfy the relation
\be \label{const_omega}
2 \ii \phi_e - \ii \Omega -1= \pm 2 \pi T \,,
\ee 
where $T= 1/\beta$. At this point one can then introduce the replica parameter by imposing $T = 1/(2\pi n)$, and write the on-shell action in terms of two out of the three parameters in \eqref{const_omega}, achieving a generalization of the SRE.
A field theory computation, starting for instance from the results in \cite{Murata:2008bg}, is still unknown. Notice that \eqref{const_omega} is the generalization of eq. \eqref{constr_chempot} to the presence of chemical potential for angular momentum. Defining the shifted potentials, whose meaning will be clear in a moment, 
\bea
\varphi \equiv \beta (\phi_e - \phi_*) \, , \qquad
\omega \equiv \beta(\Omega - \Omega_*) \, ,
\eea
where $\Omega_* = -\ii$, $\phi_* = -\ii$, we are able to write \eqref{onsh-simpl} as 
\be \label{onshell_simple}
I=  \ii \frac{\varphi^2}{2 \omega} \, .
\ee
The variables $\varphi$ and $\omega$ satisfy $ 2 \varphi -\omega = \mp 2 \pi \ii $, like in \cite{Choi:2018fdc}.  This redefinition is similar in spirit to that performed in \cite{Cabo-Bizet:2018ehj,Cassani:2019mms}, where $\Omega_*$ and $\phi_*$ are the values of the chemical potentials computed on the extremal BPS solution. In our case indeed these values give $T=0$, however the corresponding horizon radius $r_*$ is imaginary, $r_* =\pm \ii \sqrt{j}$. While it is hard to make sense of this as a proper ``extremal BPS" limit, the similarity that arises with \cite{Cabo-Bizet:2018ehj,Cassani:2019mms} is suggestive ($r_*$ in these latter papers is real and corresponds to a well-defined extremal BPS black hole). As a final remark, the form of the on-shell action \eqref{onshell_simple} is compatible with the more general form 
\be\label{actQis}
I= \ii \frac{\sqrt{\varphi_1 \varphi_2 \varphi_3 \varphi_4} }{2\omega} \, , \qquad \sum_{I=1}^4 \frac{\varphi_i}{2} - \omega = - 2\pi \ii \, ,
\ee
expected from the study of \cite{Choi:2018fdc} carried out for black holes with a spherical horizon. Indeed, \eqref{actQis} reduces to \eqref{onshell_simple} if we set all $\varphi_I$ equal, as is the case for minimal gauged supergravity.

\bibliographystyle{ytphys}

\bibliography{SRE-HTY}

\end{document}